\input harvmac
\def\MM{{\cal M}}
\font\zfont = cmss10 
\font\zfonteight = cmss8 
\def\ZZ{\hbox{\zfont Z\kern-.4emZ}}
\def\ZZs{\hbox{\zfonteight Z\kern-.4emZ}}

\rightline{EFI-98-58, ILL-(TH)-98-06}
\Title{
\rightline{hep-th/9812027}
}
{\vbox{\centerline{String Theory in Magnetic Monopole Backgrounds}}}
\medskip
\centerline{\it
David Kutasov${}^{1}$, Finn Larsen${}^{1}$, and Robert G. Leigh${}^{2}$
}
\bigskip
\centerline{${}^1$Enrico Fermi Institute, University of Chicago,
5640 S. Ellis Av., Chicago, IL 60637}
\centerline{${}^2$Department of Physics, University of Illinois at
Urbana-Champaign, Urbana, IL 61801}
\smallskip

\vglue .3cm
\bigskip

\noindent
We discuss string propagation in the near-horizon geometry generated
by Neveu-Schwarz fivebranes, Kaluza-Klein monopoles and fundamental
strings. When the fivebranes and KK monopoles are wrapped around a 
compact four-manifold $\MM$, the geometry is 
$AdS_3\times S^3/\ZZ_N\times \MM$ and the spacetime dynamics is 
expected to correspond to a local two dimensional conformal field 
theory. We determine the moduli space of spacetime CFT's, study the 
spectrum of the theory and compare the chiral primary operators 
obtained in string theory to supergravity expectations.

\Date{12/98}

\def\journal#1&#2(#3){\unskip, \sl #1\ \bf #2 \rm(19#3) }
\def\andjournal#1&#2(#3){\sl #1~\bf #2 \rm (19#3) }

\def\hat{\widehat}
\def\ie{{\it i.e.}}
\def\eg{{\it e.g.}}

\def\frac#1#2{{#1\over#2}}

\def\inbar{\,\vrule height1.5ex width.4pt depth0pt}
\def\IC{\relax\hbox{$\inbar\kern-.3em{\rm C}$}}
\def\IR{\relax{\rm I\kern-.18em R}}
\def\IP{\relax{\rm I\kern-.18em P}}

%
%
\def\np#1#2#3{Nucl. Phys. {\bf B#1} (#2) #3}

\def\plb#1#2#3{Phys. Lett. {\bf #1B} (#2) #3}
\def\prl#1#2#3{Phys. Rev. Lett. {\bf #1} (#2) #3}

\def\prd#1#2#3{Phys. Rev. {\bf D#1} (#2) #3}

\def\cmp#1#2#3{Comm. Math. Phys. {\bf #1} (#2) #3}

\def\JHEP#1#2#3{J. High Energy Phys. {\bf #1} (#2) #3}
\def\atmp#1#2#3{Adv.Theor.Math.Phys. {\bf #1} (#2) #3}

\catcode`\@=11
\def\slash#1{\mathord{\mathpalette\c@ncel{#1}}}
\overfullrule=0pt

\def\CC{{\cal C}}

\def\LL{{\cal L}}
\def\NN{{\cal N}}

\def\VV{{\cal V}}
\def\WW{{\cal W}}
\def\XX{{\cal X}}
\def\YY{{\cal Y}}

\def\underrel#1\over#2{\mathrel{\mathop{\kern\z@#1}\limits_{#2}}}

\catcode`\@=12


%

\def\exp{{\rm exp}}



\newsec{Introduction}
\lref\chs{C.C. Callan, J. Harvey, and A. Strominger, hep-th/9112030, 
in Proceedings, {\it String Theory and Quantum Gravity}, Trieste, 1991.}%
\lref\seibergns{N. Seiberg, hep-th/9705221, \plb{408}{1997}{98}.}%
\lref\juanads{J. Maldacena, hep-th/9711200, \atmp{2}{1998}{231}.}%
\lref\abks{O. Aharony, M. Berkooz, D. Kutasov, and 
N. Seiberg, hep-th/9808149.}%
\lref\gks{A. Giveon, D. Kutasov, and N. Seiberg, hep-th/9806194.}%
\lref\intril{K. Intriligator, hep-th/9708117, \atmp{1}{1998}{271}.}%
\lref\mmmref{
Y-K. Cheung and Z. Yin, hep-th/9710206, \np{517}{1998}{69};
R. Minasian and G. Moore, hep-th/9710230, \JHEP{11}{1997}{002};
J. Maldacena, A. Strominger and E. Witten, hep-th/9711053,
\JHEP{12}{1997}{002}; 
D. Freed, J. Harvey, R. Minasian, and G. Moore, hep-th/9803205,
\atmp{2}{1998}{601}.}%
\lref\gps{S. Giddings, J. Polchinski, and A. Strominger, hep-th/9305083,
\prd{48}{1993}{5784}.}%
\lref\johnson{C. Johnson, hep-th/9403192, \prd{50}{1994}{4032};
hep-th/9406069, \prd{50}{1994}{6512}.}%
\lref\cftcydyon{M. Cveti\v{c} and A. Tseytlin, hep-th/9510097, 
\plb{366}{1996}{95};
hep-th/9512031, \prd{53}{1997}{5619};
K. Behrndt, I. Brunner, and I. Gaida, hep-th/9804159, 
\plb{432}{1998}{310};
hep-th/9806195.}
\lref\bfss{T. Banks, W. Fischler, S. Shenker, and L. Susskind, 
hep-th/9610043, \prd{55}{1997}{5112}.}
\lref\polyads{S. Gubser, I. Klebanov, and A. Polyakov, hep-th/9802109,
\plb{428}{1998}{104}}
\lref\wittads{E. Witten, hep-th/9802150, \atmp{2}{1998}{253}.}

It is currently believed that many (perhaps all) vacua of string theory
have the property that their spacetime dynamics can be alternatively
described by a theory without 
gravity~\refs{\bfss,\seibergns,\juanads,\abks}. 
This theory is in general non-local, 
but in certain special cases 
it is expected to become a local quantum field theory (QFT). It is 
surprising that string dynamics can be equivalent to a local QFT. A better
understanding of this equivalence would have numerous applications
to strongly coupled gauge theory, black hole physics and a 
non-perturbative
formulation of string theory.

An important class of examples for which string dynamics is described  
by local QFT is string propagation on manifolds that include an 
anti-de-Sitter spacetime $AdS_{p+1}$\refs{\juanads,\polyads,\wittads}. 
In this case the corresponding theory without gravity is a $p$ 
dimensional conformal field theory (CFT). In general, solving the string 
equations of motion on $AdS_{p+1}$ requires turning on Ramond-Ramond 
(RR) backgrounds, which are not well understood. This makes it difficult 
to study  the $AdS/CFT$ correspondence in string theory
and most of the work on this subject is restricted to situations
where the curvature on $AdS_{p+1}$ is small and the low energy
supergravity approximation is reliable.  
 
\lref\amit{S. Elitzur, O. Feinerman,
A. Giveon and D. Tsabar, hep-th/9811245.}%
String theory on $AdS_3$ is special in several respects. First, 
in this case the ``dual'' CFT is two dimensional and the corresponding
conformal symmetry is infinite dimensional. In general,
two dimensional CFT's are better understood than their higher
dimensional analogs and one may hope that this will also be the case
here. Second, string theory on $AdS_3$ can be defined without
turning on RR fields and thus should be more amenable to traditional
worldsheet methods. Perturbative string theory on $AdS_3$ was studied 
in \gks. The purpose of this paper is to continue this study and to apply 
it to some additional examples that are of interest in the different 
contexts mentioned above. Another application of the results of
\gks\ appears in \amit.

In section 2 we introduce the brane configuration whose near-horizon
geometry will serve as the background for string propagation later
in the paper. The configuration of interest includes $NS5$-branes,
Kaluza-Klein (KK) monopoles and fundamental strings.
We describe the supergravity solution and its near-horizon limit, and 
review some earlier results on chiral primary operators that are visible 
in supergravity. We also determine the moduli space of vacua in this 
geometry; by the $AdS/CFT$ correspondence this gives the moduli space 
of dual CFT's. The resulting moduli space, which can be
thought of as the moduli space of M-theory on $AdS_3\times
S^2\times T^6$, is given in (2.17). The duality group is $F_{4(4)}(
\ZZ)$, a discrete, non-compact version of the exceptional group $F_4$.
 
In section 3 we review the work of \gks\ on string theory on $AdS_3
\times S^3\times T^4$, and elaborate on some aspects of it. We find the
spectrum of chiral primaries in string theory and compare it to
supergravity. We also comment on a proposed identification of string
theory on $AdS_3\times S^3\times\MM$ with CFT on the symmetric product
$\MM^n/S_n$, and show that the spectra of $U(1)^4$ affine Lie algebras
in string theory and in CFT on the symmetric product disagree. We also
briefly describe the extension of the work of \gks\ to heterotic string 
theory and show that the resulting spacetime CFT is ``heterotic'' as well
(\ie\ its left and right central charges are different).  

In sections 4, 5 we discuss string propagation in the magnetic monopole 
background of section 2. The near-horizon geometry includes a Lens space 
$S^3/\ZZ_N$; therefore we start in section 4 with a description of CFT on 
Lens spaces. In section 5 we turn to string theory on such spaces and discuss 
in turn bosonic, heterotic and type II strings. We discuss the spectrum,
obtain the moduli space of vacua, and compare the resulting structure to 
the supergravity analysis of section 2. We show that the set of chiral 
operators in string theory on $AdS_3\times S^3/\ZZ_N\times T^4$ is
much larger than that in the corresponding low energy supergravity
theory. In particular, it includes an exponentially large  density
of perturbative string states which carry momentum and winding around
an $S^1$ in $S^3/\ZZ_N$. 
We also show that string theory in a monopole background exhibits an
effect familiar from quantum mechanics, the shift of angular momenta 
of electrically charged particles in the background of a magnetic 
monopole. Two appendices contain conventions, results and derivations 
used in the text. 

\newsec{Supergravity Analysis}

\lref\dyon{M. Cveti\v{c} and D. Youm, hep-th/9507090, 
\prd{53}{1996}{584}.}%
\lref\mmm{I. Klebanov and A. Tseytlin, hep-th/9604166,
\np{475}{1996}{179}; V. Balasubramanian and F. Larsen, 
hep-th/9604189, \np{478}{1996}{199}.}%

\subsec{Brane Configuration}

Consider M-theory compactified on a six dimensional manifold
$\NN$ parametrized by $(x^4, x^5, x^6, x^7, x^8, x^{11})$, 
down to five non-compact dimensions $(x^0, x^1, x^2, x^3, x^9)$. 
We will concentrate  
on the case $\NN=T^6$, but will comment briefly on
the cases where $\NN$ is $K3\times T^2$ or a Calabi-Yau
manifold. Since we would like to use weak coupling techniques, 
we identify $x^{11}$ with the M-theory direction, and send its 
radius to zero. In the resulting weakly coupled string theory 
we consider the following brane configuration~\refs{\cftcydyon}:
\item{(a)}
$N$ KK monopoles wrapped around the $T^4$ labeled by
$(x^5,\cdots, x^8)$, infinitely extended in $x^9$
 and charged under the gauge field 
$A_\mu=G_{\mu4}$ $(\mu=0,1,2,3)$.  
\item{(b)}
$N'$ $NS5$-branes wrapped around the above $T^4$
and extended in $x^9$. 
\item{(c)}
$p$ fundamental strings infinitely stretched in $x^9$. 

\noindent
The configuration (a) -- (c) preserves four supercharges
which form a chiral $(4,0)$ supersymmetry algebra in
the $1+1$ dimensional non-compact spacetime $(x^0, x^9)$
shared by all the branes. It will be useful later to note
that all the unbroken supercharges originate from the same 
worldsheet chirality. At low energies the theory on the branes 
decouples from bulk string dynamics and approaches a $(4,0)$ 
superconformal field theory. 

M-theory on $T^6$ has a large U-duality group which can
be used to relate the above brane configuration to many others,
such as that of three $M5$-branes wrapped around 
different four-cycles in $T^6$ and intersecting along the $x^9$
direction \mmm. The specific realization (a) -- (c) is 
special in that only Neveu-Schwarz (NS) sector fields are
excited; therefore we will be able to use the results of
\gks\ to study the near-horizon dynamics. 

The classical supergravity fields around the above collection
of $NS5$-branes, KK monopoles
and fundamental strings are as follows. The string frame metric is
\eqn\cymetric{\eqalign{
ds^2 =&
H_5\left[ H_{K}^{-1}\left(dx_4 + P_{K}(1-\cos\theta)d\phi\right)^2
+ H_{K}\left(dr^2 + r^2(d\theta^2+ \sin^2\theta d\phi^2)\right)
\right] + \cr
+& F^{-1}(-dt^2+dx_9^2)+\sum_{i=5}^8 dx^2_i
}}
where the harmonic functions
\eqn{\harm}{\eqalign{
H_5 =& 1 + {P_5\over r}\cr
H_{K} =& 1 + {P_{K}\over r}\cr
F =& 1 + {Q\over r}
}}
are associated with $NS5$-branes, KK monopoles, and 
fundamental strings, respectively, and we have parametrized
the space transverse to the branes $(x^1, x^2, x^3)$ by
spherical coordinates $(r,\theta, \phi)$. The dilaton and
NS $B_{\mu\nu}$ field are: 
\eqn{\harmo}{\eqalign{
B_{t9} =& ~F \cr
B_{\phi 4} =& ~P_5(1-\cos\theta)\cr
e^{-2[\Phi_{10}(r)-\Phi_{10}(\infty)]} =& ~{F\over H_5}\cr
}}
The charges  $P_5$, $P_{K}$, $Q$ in \harm\ are related to the 
numbers of branes $N'$, $N$ and $p$ via:
\eqn{\quant}{\eqalign{
P_5 = & ~{\alpha^\prime\over 2R} ~N' \cr
P_{K} = &~ {R\over 2}~N\cr
Q = & ~{\alpha^{\prime 3}g^2_{\rm s}\over 2RV} ~p 
}}
where $R$ is the radius of $x^4$ (asymptotically far from the branes),
$g_s\equiv \exp\Phi_{10}(\infty)$
and $(2\pi)^4V$ is the volume of the $T^4$. Note that:
\item{(a)} From the point of view of the $3+1$ dimensional
non-compact spacetime labeled by $x^\mu$, 
$\mu=0,1,2,3$, the vacuum \cymetric\ 
-- \quant\ is magnetically charged under two gauge fields,
$G_{\mu4}$, $B_{\mu4}$. The magnetic charges are $N$ and $N'$,
respectively. 
\item{(b)} One can make the vacuum \cymetric\ -- \quant\ arbitrarily
weakly coupled everywhere by sending $g_s\to0$ and $p\to\infty$
with $Q$ fixed.
\item{(c)} T-duality in $x^4$ exchanges $NS5$-branes and KK 
monopoles. Thus it exchanges $N$ and $N'$ as well as type IIA
and IIB.
\item{(d)}  
The fields that are excited in \cymetric\ -- \quant\ exist in 
all closed string theories, including the bosonic string, 
the heterotic string, and the type II superstring. Therefore,
one can study this solution in all these theories. 
 
\subsec{The near-horizon limit}

A dual description of the decoupled low energy dynamics on the branes 
is obtained by studying string dynamics in the background 
\cymetric\ -- \quant\ in the near-horizon limit $r\to 0$ \juanads.  
In this limit the metric reduces to\foot{Many similar examples are
discussed in 
\ref\lens{H. Boonstra, B. Peeters and K. Skenderis, hep-th/9803231,
\np{533}{1998}{127}; M. Cveti\v{c}, H. Lu, and C.N. Pope, hep-th/9811107.}
.}:
\eqn{\nearh}{\eqalign{
ds^2=& {P_5\over P_{K}}\left[dx_4 + P_{K}(1-\cos\theta)d\phi\right]^2 
+ P_5 P_{K}(d\theta^2+ \sin^2\theta d\phi^2)
\cr
+&  {P_5 P_{K}\over r^2}dr^2+ {r\over Q}(-dt^2+dx_9^2)+\sum_{i=5}^8 dx^2_i
}}
The $B_{\mu\nu}$ field is given by \harmo. The dilaton, which far from
the branes is arbitrary, becomes a fixed scalar in the near-horizon
geometry \nearh:
\eqn\fixdil{e^{-2\Phi^{\rm hor}_6}=e^{-2\Phi_{\rm 10}}{V\over
\alpha^{\prime 2}} = {Q\over P_5}~{V\over\alpha^{\prime 2}}
~e^{-2\Phi_{10\infty}} = {p\over N'}
}
The three dimensional 
space parametrized by $t$, $r$, and $x^9$ becomes
after the coordinate change $\rho^2 = 4P_5P_{K}r/Q$: 
\eqn\ads{
ds^2 = {l^2\over \rho^2}d\rho^2 +  {\rho^2\over l^2}(-dt^2 + dx_9^2) }
where 
\eqn\lsqu{l^2 = 4P_5P_{K}=\alpha'NN'}
The metric \ads\ is that of $AdS_3$
with curvature $\Lambda= - 1/l^2$. 

We next turn to the three dimensional space parametrized 
by $\theta$, $\phi$, and $x_4$.
Far from the
branes, the radius of $x^4$, $R$, is a free parameter, the expectation
value of a massless scalar field. In the near-horizon geometry 
\nearh\ $R$ is fixed:
\eqn\fixx{ R^{\rm hor}_4 = \sqrt{G_{44}}~R = R\sqrt{P_5\over P_{K}}
= \sqrt{N'\alpha^\prime\over N}}
The corresponding scalar field is massive. 
The radius of $x^4$ \fixx\ is typically of string size (if $N$ and 
$N'$ are comparable). Hence, at low energies we can dimensionally 
reduce along $x^4$. This leaves an $S^2$ labeled by $\theta$, $\phi$, 
which can be interpreted as the horizon of a four dimensional 
black hole, or a five dimensional black string. In the full theory 
$x^4$ is retained and the resulting three dimensional space can be 
identified with the manifold $S^3/\ZZ_N$. Altogether, we are led to 
study string theory on 
\eqn\backgg{AdS_3\times S^3/\ZZ_N\times T^4} 

Another difference between the asymptotic and near-horizon geometries
is that while, as we saw, the asymptotic background \cymetric\ --
\quant\ is magnetically charged under both $G_{\mu4}$ and $B_{\mu4}$
(with different magnetic charges $N$, $N'$), in the near-horizon
geometry \nearh\ 
$G_{\phi 4}-B_{\phi 4}=0$. Thus, the vacuum \nearh\ is magnetically
charged under $G_{\mu4}+B_{\mu4}$ only, 
\eqn{\fixu}{G_{\phi 4}+B_{\phi 4}=2P_5 ( 1- \cos\theta )}
Recalling that $G+B$ and $G-B$ couple to different chiralities on the
worldsheet, we see that the background in question is $S^2\times S^1$
with the $SO(3)$ isometry of the two-sphere arising from the worldsheet
chirality\foot{The foregoing discussion is valid for large $N$, $N'$,
when supergravity is reliable. We will see later that when $N$ or $N'$
are $\le 2$, a second $SO(3)$ arises from the other worldsheet
chirality.} that does not couple to $G+B$, which we will refer to
as right-moving. We conclude that the $\ZZ_N$ orbifold in \backgg\ is
asymmetric.

\subsec{Black Holes and Spacetime CFT}

One of the motivations for the present work is the relation to black 
holes. The configuration given in \cymetric\  
can be generalized by adding one more charge, corresponding to momentum 
along the 9th direction. If $x^9$ is compact, the 
resulting metric is that of a regular black hole in four 
dimensions, the  Cveti\v{c}-Youm dyon \refs{\dyon}. 
Such black holes can be interpreted as excitations of the configuration 
\cymetric. The black hole entropy follows from the
large degeneracy of these excitations. Understanding the
spacetime dynamics of strings on \backgg\ involves understanding
precisely these excitations, and thus is important for black hole 
physics. 

\lref\brhen{J. D. Brown and M. Henneaux, 
\cmp{104}{1986}{207}.}%
\lref\strom{A. Strominger, hep-th/9712251,
\JHEP{02}{1998}{009}.}%
Brown and Henneaux \brhen\ 
have shown that any theory of gravity
on $AdS_3$ has a large symmetry algebra containing two copies
of the Virasoro algebra with central charge 
\eqn\cspacetime{c_{\rm st}\simeq {3l\over 2 l_p}}
where $l$ is the radius of curvature of $AdS_3$ \lsqu, 
and $l_p$ is the three dimensional
Newton constant. The calculation of \brhen\ is semiclassical 
and is expected in general to receive both quantum gravity corrections
which are suppressed by powers of $l_p/l$, and string corrections
suppressed by powers of $l_s/l$ ($l_s^2\equiv\alpha'$). 
We will see examples of such
corrections below. In our case, the semiclassical computation
gives 
\eqn\cssp{c_{\rm st}=6NN'p}  
Strominger \strom\ pointed out that if the spacetime dynamics
of a theory of gravity on $AdS_3$
corresponds to a unitary and modular invariant 
CFT with the central charge \cspacetime\ (and satisfies
certain additional mild assumptions), then the
standard CFT degeneracy of states agrees with the  
Bekenstein-Hawking entropy of the corresponding black holes.
For the case of interest here the details were discussed in 
\ref\bl98{V. Balasubramanian and F. Larsen, 
hep-th/9802198, \np{528}{1998}{229}.}.
Thus in our study of string theory on \backgg\ 
we will be interested in computing the central charge 
\cssp\ and any stringy corrections to it, and understanding
the spectrum of operators that contribute to the
BH\foot{$=$ Black Hole, Bekenstein-Hawking, Brown-Henneaux}
entropy.

\subsec{The Spectrum of Perturbations}
\lref\flspectrum{F. Larsen, hep-th/9805208.}
\lref\dbspectrum{J. de Boer, hep-th/9806104.}

Later we will study the spectrum of perturbations of
string theory in the background \backgg. Some of these
perturbations are visible already in the supergravity
approximation. In this subsection we summarize the results
of \refs{\flspectrum, \dbspectrum} regarding the spectrum
of chiral operators in supergravity.

Consider eleven dimensional supergravity compactified on
\eqn\comel{AdS_3\times S^2\times \NN}
where $\NN$ is a Calabi-Yau manifold, $K3\times T^2$ or 
$T^6$. We will assume that the manifold $\NN$ has Planck 
(or string) scale 
size, and dimensionally reduce all supergravity fields on it.
The size of $S^2$ is assumed to be large; 
therefore we will keep Kaluza -- Klein harmonics on 
the sphere.
The theory preserves $(4,0)$ supersymmetry and, as explained 
in the previous subsection, is equivalent to a two dimensional
CFT. 

States are labeled by the quantum
numbers $h$, $\bar h$, $\bar j$, where $h$ is the scaling
dimension with respect to the left-moving Virasoro 
algebra of \brhen, while $\bar h$ and $\bar j$
are the quantum numbers under the right-moving
$N=4$ superconformal algebra
(\ie\ under right-moving 
Virasoro and $SU(2)_R$). Note that an
$SL(2)_R\times SL(2)_L$ subalgebra of Virasoro is identified
with the isometry of $AdS_3$ in \comel\ while the $SU(2)_R$
symmetry is the $SO(3)$ isometry of the two-sphere.

Reducing the eleven dimensional supergravity fields on the manifold
\comel\ gives rise to short representations of the $N=4$ superconformal 
algebra. To analyze the resulting spectrum it is convenient
to perform the reduction in two steps, first reducing to five dimensions
on the manifold $\NN$, and then further reducing on
$AdS_3\times S^2$. After reducing on $\NN$, one finds the
following spectrum:
\item{(a)} A gravity multiplet, whose bosonic components are
the five dimensional graviton and a gauge field (the graviphoton).
This multiplet contains eight bosonic and eight fermionic degrees
of freedom.
\item{(b)} 
$n_H$ hypermultiplets, consisting of two real scalars
and fermions. Each multiplet thus has two bosonic and two
fermionic degrees of freedom. 
\item{(c)} $n_V$ vectormultiplets consisting of a vector field,
a scalar and fermions ($4+4$ physical degrees of freedom). 
\item{(d)} $n_S$ gravitino multiplets consisting of two vectors
and fermions ($6+6$ degrees of freedom).  
 
\noindent
The values of $n_H$, $n_V$, and $n_S$ for 
different choices of the manifold $\NN$ are
\ref\cyspectrum{A. Cadavid, A. Ceresole, R. D'Auria, and S. Ferrara, 
hep-th/9506144, \plb{357}{1995}{76};
G. Papadopoulos and P. Townsend, hep-th/9506150,
\plb{357}{1995}{300}.}:
\item{(a)} On a Calabi-Yau manifold, at a generic point in moduli 
space, $n_H = 2(h_{21}+1)$, $n_V = h_{12}-1$, and $n_S=0$.
\item{(b)} On $K3\times T^2$, generically $n_V=22$, $n_H = 2(n_V-1)$,
and $n_S=2$. At points in moduli space where the gauge symmetry is 
enhanced, $n_V$ increases accordingly. Note that this problem
is dual to the heterotic string on a torus, a case we will discuss 
below. 
\item{(c)} On $T^6$, $n_H=n_V=14$ and $n_S=6$.

\noindent
The further reduction on $AdS_3\times S^2$ gives the following spectrum 
of 
chiral primaries:
\bigskip
\vbox{
$$\vbox{\offinterlineskip
\hrule height 1.1pt
\halign{&\vrule width 1.1pt#
&\strut\quad#\hfil\quad&
\vrule width 1.1pt#
&\strut\quad#\hfil\quad&
\vrule width 1.1pt#
&\strut\quad#\hfil\quad&
\vrule width 1.1pt#\cr
height3pt
&\omit&
&\omit&
&\omit&
\cr
&\hfil $h-\bar h$&
&\hfil degeneracy&
&\hfil range of $\bar h=\bar j$&
\cr
height3pt
&\omit&
&\omit&
&\omit&
\cr
\noalign{\hrule height 1.1pt}
height3pt
&\omit&
&\omit&
&\omit&
\cr
&\hfil $1/2$&
&\hfil $n_H$&
&\hfil $1/2,3/2,\ldots$&
\cr
height3pt
&\omit&
&\omit&
&\omit&
\cr
\noalign{\hrule}
height3pt
&\omit&
&\omit&
&\omit&
\cr
&\hfil $0$&
&\hfil $n_V$&
&\hfil $1,2,\ldots$&
\cr
\noalign{\hrule}
height3pt
&\omit&
&\omit&
&\omit&
\cr
&\hfil $1$&
&\hfil $n_V$&
&\hfil $1,2,\ldots$&
\cr
\noalign{\hrule}
height3pt
&\omit&
&\omit&
&\omit&
\cr
&\hfil $-1/2$&
&\hfil $n_S$&
&\hfil $3/2,5/2,\ldots$&
\cr
\noalign{\hrule}
height3pt
&\omit&
&\omit&
&\omit&
\cr
&\hfil $1/2$&
&\hfil $n_S$&
&\hfil $3/2,5/2,\ldots$&
\cr
\noalign{\hrule}
height3pt
&\omit&
&\omit&
&\omit&
\cr
&\hfil $3/2$&
&\hfil $n_S$&
&\hfil $1/2,3/2,\ldots$&
\cr
\noalign{\hrule}
height3pt
&\omit&
&\omit&
&\omit&
\cr
&\hfil $-1$&
&\hfil $1$&
&\hfil $2,3,\ldots$&
\cr
\noalign{\hrule}
height3pt
&\omit&
&\omit&
&\omit&
\cr
&\hfil $0$&
&\hfil $1$&
&\hfil $2,3,\ldots$&
\cr
\noalign{\hrule}
height3pt
&\omit&
&\omit&
&\omit&
\cr
&\hfil $1$&
&\hfil $1$&
&\hfil $1,2,\ldots$&
\cr
\noalign{\hrule}
height3pt
&\omit&
&\omit&
&\omit&
\cr
&\hfil $2$&
&\hfil $1$&
&\hfil $1,2,\ldots$&
\cr
height3pt
&\omit&
&\omit&
&\omit&
\cr
}\hrule height 1.1pt
}
$$
}
\centerline{\sl Table 1: Spectrum of chiral primaries for $AdS_3\times 
S^2\times {\cal N}$.}
\def\sugrastates{Table 1}
\noindent In section 5 we will find a stringy generalization of the 
spectrum of
\sugrastates\ for toroidally compactified heterotic and type II 
string theories, and will see that the above table is indeed obtained 
in the supergravity limit.

\subsec{Moduli Space of Vacua}

The purpose of this subsection is to determine 
the moduli space of vacua of M-theory 
on the manifold \comel\ for the case $\NN=T^6$.
The moduli space of M-theory on $T^6\times \IR^{4,1}$
is
\eqn\modesix{
E_{6(6)}(\ZZ)  \backslash E_{6(6)}/ USp(8)
}
$E_{6(6)}$ is a non-compact form of $E_6$ with maximal 
compact subgroup $USp(8)$. Black strings 
in the five non-compact dimensions are charged under 
the various $B_{\mu\nu}$ fields (which in five dimensions
are dual to gauge fields). 
There are $27$ independent strings transforming in the ${\bf 27}$
of the $E_{6(6)}(\ZZ)$ U-duality group \modesix: 
the $M5$-branes with four of their 
dimensions wrapped on a $T^4$ (15 possibilities), the $M2$-branes 
with one dimension wrapped on an $S^1$ (6 of these), and the KK monopoles
charged under the six gauge fields $G_{\mu,i}$
and wrapped around the remaining $T^5$. It
is convenient to organize these 27 charges into an $8\times 8$
symplectic-traceless antisymmetric matrix (utilizing the
maximal $USp(8)$ subgroup of $E_6$):
\eqn{\esixspeight}{
	{\bf 27}=\pmatrix{b\ J_{(1)} & Z\cr
	-Z^T & -{1\over3}b\ J_{(3)}+A_{ij}T^{ij}}
	}
where $J_{(i)}$ are the symplectic invariants of $USp(2i)$, and
$T^{ij}$ are a basis of traceless antisymmetric $6\times 6$ matrices.
One can choose 
the $6\times 2$ charges $Z$ in \esixspeight\ to correspond to the
$M2$-brane and KK monopole charges described above, while $A_{ij}$
and $b$ parametrize the $M5$-brane charges. 

Eq. \esixspeight\ makes manifest the decomposition
of the ${\bf 27}$ of $E_6$ in terms of representations of its
$USp(2)\times USp(6)$ subgroup: $b{\bf(1,1)}+Z{\bf (2,6)}+A{\bf (1,14)}$. 
The near-horizon geometry \nearh\ 
is U-dual to a configuration of three $M5$-branes with charges:
$\langle b\rangle\neq 0$, $\langle Z\rangle=\langle A_{ij}\rangle=0$
\flspectrum. To find the subgroup of $E_{6(6)}(\ZZ)$ that is left
unbroken by $\langle b\rangle$, one notes that $E_6$ has a maximal
$F_4$ subgroup, under which the  
${\bf 27}$ decomposes 
as ${\bf 26+1}$. The $USp(2)\times USp(6)$ 
discussed above is the maximal compact subgroup of $F_4$. Therefore,
$b$ is a singlet under $F_4$, and the subgroup of $E_{6(6)}(\ZZ)$
U-duality preserved by the near-horizon geometry is $F_{4(4)}(\ZZ)$.
The moduli space is the coset:
\eqn\modspc{
F_{4(4)}(\ZZ)  \backslash F_{4(4)}/ USp(2)\times USp(6)
}
The adjoint of $F_4$, the ${\bf 52}$, decomposes
under $USp(2)\times USp(6)$
as ${\bf [(3,1)+(1,21)] + (2,14)}$. The noncompact form thus has signature
$(28,24)$, and the coset has dimension 28. The $2\times 14$ moduli 
correspond
to the $n_H=14$ hypermultiplets with $h=1$, $\bar h=1/2$ in \sugrastates.
In Section 5 we will reproduce aspects of the moduli space \modspc\ in 
string theory.

\newsec{String Theory on $AdS_3\times S^3\times T^4$}

The near-horizon geometry of a system of $k$ $NS5$-branes
and $p$ fundamental strings in type II string theory on
a four-manifold $\MM$ ($\MM=T^4$ or $K3$) is 
\eqn\adst{AdS_3\times S^3\times \MM}
Type II string theory on the manifold \adst\ has 
$(4,4)$ superconformal symmetry when $\MM=T^4$ and
in the type IIB case also when $\MM=K3$. It has $(4,0)$
superconformal symmetry for the 
heterotic string on $\MM=T^4$, $K3$, 
and for type IIA on $\MM=K3$ (which is dual to the
heterotic theory on $T^4$). String theory on manifolds
including an $AdS_3$ factor was discussed in \gks, and the case
of type II string propagation  
on \adst\ with $\MM=T^4$ was described in detail. 
In this section we will review this construction, as a warmup
for the case \backgg. We 
will also make some comments on the spectrum of the theory, and briefly 
discuss the heterotic case.
Some of the
conventions and other details appear in Appendix A.  

\subsec{Symmetries of string theory on $AdS_3\times S^3\times T^4$}

The worldsheet and spacetime symmetries of
string theory on \adst\ act separately  
on the left and right-movers on the worldsheet. 
Therefore in this subsection we will discuss
the chiral (holomorphic) symmetry structure,
both on the worldsheet and in spacetime. 
This will also be useful when we turn to string
theory on \backgg\ whose (anti-) holomorphic structure
is identical to that discussed here. 

The affine worldsheet symmetry of 
string theory on $AdS_3\times S^3\times T^4$
is $\widehat{SL(2)}\times \widehat{SU(2)}\times \widehat{U(1)}^4$. 
It is realized as follows. There 
are three bosonic currents $j^A$, realizing 
$\widehat{SL(2)}$ at level $k+2$, and three fermions $\psi^A$, 
forming an $\widehat{SL(2)}$ at level $-2$. 
Similarly, there are three bosonic currents $K^a$, realizing an 
$\widehat{SU(2)}$ at level $k-2$, 
and three fermions $\chi^a$, giving an 
$\widehat{SU(2)}$ at level 2. The total currents
\eqn\Jp{\eqalign{
J^A=&~j^A-{i\over k}\epsilon^A_{\,\,\,BC}\psi^B\psi^C,\qquad A,B=+,-,3\cr
K^a=&~k^a-{i\over k}\epsilon^a_{\,\,\,bc}\chi^b\chi^c,~~~\qquad 
~~~a,b=+,-,3\cr
}}
thus have the same level $k$. The total worldsheet central charge of the
$SL(2)\times SU(2)$ theory is identical to its flat space value ($c=9$), 
for all values of $k$.

The $\widehat{U(1)}^4$ is realized in terms of 
free bosonic currents $i\partial Y^j$ and free fermions $\lambda^j$, 
$j=1,2,3,4$. The worldsheet theory is superconformal; 
the supercurrent is
\eqn\gsl{T_F(z)={2\over k}(\eta_{AB}\psi^Aj^B
-{i\over 3k}\epsilon_{ABC}\psi^A\psi^B\psi^C)
+{2\over k}(\chi^ak_a
-{i\over 3k}\epsilon_{abc}\chi^a\chi^b\chi^c)+\lambda^i \partial Y_i}
Generally in string theory affine symmetries on the worldsheet give rise 
to gauge symmetries in spacetime. Contour integrals of the worldsheet 
generators give global charges, corresponding to gauge transformations
that do not vanish rapidly enough at infinity. The analysis of 
\brhen\ shows that in gravity on $AdS_3$ one should allow a rich set 
of non-trivial behaviors of the metric at infinity. This leads to  
a large spacetime global symmetry group, the $2d$ conformal group. 
Similarly, allowing non-trivial behaviors of gauge fields 
at the boundary of $AdS_3$
leads to an enhancement of global spacetime
gauge symmetries to the corresponding
affine Lie algebras \gks.  

\lref\wakim{M. Wakimoto, \cmp{104}{1986}{605}.}
To obtain a worldsheet description 
of the above infinite spacetime symmetry
algebras it is convenient to use 
the Wakimoto representation of
the conformal $\sigma$-model on $AdS_3$ \wakim. This representation,
which is summarized in Appendix A, parametrizes the manifold by
the coordinates $(\phi, \gamma, \bar\gamma)$. The radial direction
is $\phi$, with the boundary of $AdS_3$ corresponding to $\phi=\infty$.
The Wakimoto description is particularly useful for studying the
structure of string theory on $AdS_3$ in the limit $\phi\to\infty$. 
In this limit the following two important simplifications occur:  
\item{(a)} The worldsheet $\sigma$-model becomes free.
\item{(b)} The string coupling goes to zero. 

\noindent
Thus the full string
theory becomes weakly coupled both
on the worldsheet and in spacetime
in this limit, 
regardless of the fixed coupling
in the original description.  
This allows one to study all aspects of string
theory on $AdS_3$ which are observable at $\phi\to\infty$ using
free field theory on the worldsheet. This
includes the infinite spacetimes symmetries, since these 
correspond to gauge
transformations which are completely specified by the behavior
at infinity. As shown in \gks, the form of the Virasoro $(L_n)$,
$\widehat{SU(2)}$ $(T^a_n)$ and $\widehat{U(1)}^4$ $(\alpha^i_n)$  
charges as $\phi\to\infty$ is:
\eqn\stvira{\eqalign{
-L_n=&\oint dz\left[
(1-n^2)J^3\gamma^n+{n(n-1)\over2}J^-\gamma^{n+1}
+{n(n+1)\over2}J^+\gamma^{n-1}\right]\cr
T^a_n=&\oint dz \{G_{-{1/2}},\chi^a(z)\gamma^n(z)\}\cr
\alpha^i_n=&\oint dz \{G_{-{1/2}},\lambda^i(z)\gamma^n(z)\}\cr
}}
The algebra satisfied by \stvira\ is:
\eqn\virfull{\eqalign{
[L_n,L_m]=&(n-m)L_{n+m}+{c_{\rm st}\over12}(n^3-n)\delta_{n+m,0}\cr
[T^a_n,T^b_m]=&i\epsilon^{ab}_{\,\,\,\,\,\,c}T^c_{n+m}+{k_{\rm st}\over2}
n\delta^{ab}\delta_{n+m,0}\cr
[L_m,T^a_n]=&-nT^a_{n+m}\cr
[\alpha_n^i,\alpha_m^j]=&pn\delta^{ij}\delta_{n+m,0}\cr
[L_m,\alpha^i_n]=&-n\alpha^i_{n+m}\cr
}}
where 
\eqn\centchgs{c_{\rm st}=6kp;\;\;\; k_{\rm st}=kp}
and $p$ is a certain winding number that characterizes the embedding
of the worldsheet into spacetime, reflecting the presence of $p$
fundamental strings in the vacuum,
\eqn\pdef{p\equiv \oint {dz\over 2\pi i}
{\partial_z\gamma(z) \over \gamma(z)}}
 
As mentioned above, string theory on \adst\ exhibits $N=4$ superconformal
invariance in spacetime. The superconformal generators are given 
by\foot{More precisely, the supercharges $Q$ form the {\it global}
$N=4$ superconformal algebra. As explained
in \gks, the full infinite superconformal symmetry can be obtained
by acting on the global supercharges with the generators \stvira. This
has been analyzed in \ref\itob{K. Ito, hep-th/9811002; 
M. Bershadsky, unpublished.}.}
\eqn\supchh{Q=\oint dz e^{-{\phi\over2}}S(z)~; \;\;\;
S(z)=e^{{i\over2}\sum_I\epsilon_I H_I(z)}}
where $H_I$, $I=1,\cdots, 5$ are the scalar fields, defined in Appendix
A, which are obtained by bosonizing the fermions,
and $\epsilon_I=\pm 1$. 
As discussed in \gks, mutual locality of the supercharges and BRST 
invariance lead to the constraints: 
\eqn\flatc{\prod_{I=1}^5\epsilon_I=
\prod_{I=1}^3\epsilon_I=1}
These projections leave eight spacetime supercharges, which together
with $L_{\pm 1}$, $L_0$ and $T^a_0$ form the global $N=4$ superconformal
algebra \gks.  

\subsec{Comments on the spectrum}

The construction of string excitations in the background 
\adst\ is very similar to that corresponding to
superstrings in
flat spacetime. The plane wave zero mode wave function familiar 
from flat spacetime is replaced by $V_{j;m,\bar m}V'_{j';m',
\bar m'}\exp(i \vec p\cdot \vec Y+
i\vec{\bar p}\cdot\vec{\bar Y})$ where $V$, $V'$ are the wave 
functions on $AdS_3$, $S^3$ respectively. Their transformation
properties under the worldsheet affine Lie algebra are described
in Appendix A. $(\vec p, \vec{\bar p})$ is a vector in an even,
self-dual Narain lattice $\Gamma^{4,4}$.
The towers of string states are obtained by
multiplying this zero mode wave function by a polynomial in
the fermionic oscillators\foot{For simplicity we restrict
here to NS-NS sector excitations. The generalization to
other sectors is straightforward. We also work at generic
values of the momenta; for special values there might be
physical states at other ghost numbers.} 
$\psi^A$, $\chi^a$, $\lambda^j$,
and the bosonic oscillators $j^A$, $k^a$, $\partial Y^j$
(and their derivatives),
and a similar polynomial in the antiholomorphic oscillators,
and restricting to the BRST cohomology. 
The most general state in the $(-1,-1)$ picture of the
NS-NS sector has the form
\eqn\nsstates{V_{NS}=e^{-\phi-\bar\phi}
P_N(\psi^A,\partial\psi^A,\cdots, j^A,\partial j^A,
\cdots)\bar
P_{\bar N}(\bar\psi^A,\cdots)V_{j;m,\bar m}V'_{j';m',\bar m'}
\exp(i \vec p\cdot \vec Y+i\vec{\bar p}\cdot\vec{\bar Y})}
where $P_N$ is a polynomial in the bosonic and fermionic 
worldsheet fields and their derivatives with scaling dimension
$N$, and similarly for $\bar P_{\bar N}$. BRST invariance
is the requirement that the matter part of \nsstates\ is
a lower component of an $N=1$ worldsheet superfield with 
$(\Delta,\bar\Delta)=({1\over2},{1\over2})$.
Thus, one must have:
\eqn\dddii{N-{j(j+1)\over k}+{j'(j'+1)\over k}
+{|\vec p|^2\over2}={1\over2}}
and a similar relation with $N\to\bar N$ and 
$\vec p\to\vec{\bar p}$ coming from the other chirality.
One can think of \dddii\ as determining $j$ (the ``energy'')
in terms of $\vec p$, $j'$ (the ``momentum'') and $N$, the
excitation level of the string.
$V_{NS}$ is null if it can be written as
the higher component of a worldsheet superfield 
with $(\Delta,\bar\Delta)=({1\over2},0)$ or $(0,{1\over2})$.

\lref\cftbook{P. Di Francesco, P. Mathieu, and  D. Senechal,
{\it Conformal Field Theory}, Springer, New York (1997).}%
The free spectrum can be summarized by writing down the 
torus partition sum. The partition sum of CFT on $AdS_3$
(more precisely on its Euclidean version $H_3^+$) appears
in \ref\gaw{K. Gawedzki, hep-th/9110076, in NATO ASI: 
Cargese 1991: 247-274.}; 
the rest is standard. Denoting the $SU(2)$ characters with spin $j$
at level $k$ by $\chi_j^{(k)}(\tau)$, the string partition
sum is ($q\equiv\exp(2\pi i\tau)$):
\eqn\partsum{\eqalign{
Z(\tau,\bar\tau)=&{1\over\sqrt{\tau_2}|\eta(\tau)|^2}
\sum_{j,\bar j\le {k\over2}-1}
\NN_{j,\bar j}\chi_j^{(k-2)}(\tau) 
\chi_{\bar j}^{(k-2)}(\bar\tau)
{1\over |\eta(\tau)|^8}
\sum_{(\vec p,\vec{\bar p})\in 
\Gamma^{4,4}}q^{{1\over2} {\vec p}^2}{\bar q}^{{1\over2} {\vec 
{\bar p}}^2}\cr
&\left|\left({\theta_3\over\eta}\right)^4-
\left({\theta_4\over\eta}\right)^4-
\left({\theta_2\over\eta}\right)^4\right|^2\cr
}}
where the matrix $\NN_{j, \bar j}$ parametrizes 
the bosonic $\widehat{SU(2)}_{k-2}$ 
modular invariant (see \eg\ \cftbook); 
we will mainly discuss the simplest case,
the $A$-series modular invariant for
which $\NN_{j,\bar j}=\delta_{j,\bar j}$.
We have factored
out an infinite overall constant from the $AdS_3$ partition sum, 
which has a clear physical meaning -- it is the infinite volume of
the $\phi$ direction (see Appendix A).

The transformation properties of physical states such as \nsstates\
under the spacetime symmetries are obtained
by computing the commutators of the generators 
\stvira, \supchh\ with the
vertices \nsstates. As an example, under the spacetime Virasoro 
algebra, the states \nsstates\ transform either as primaries or as
descendants. The primaries, $V_{\rm phys}(h;m,\bar m)$ satisfy
\eqn\primr{
[L_n, V_{\rm phys}(h;m,\bar m)]=\left(n(h-1)-m\right)
V_{\rm phys}(h;m+n,\bar m)} 
where $h$ is the scaling dimension of $V_{\rm phys}$.
A state of the form \nsstates\ is
primary under spacetime Virasoro if
it is primary (in the sense of Appendix A)
under $\widehat{SL(2)}$, however the latter is not 
a necessary condition. A large class of Virasoro
primaries that are also $\widehat{SL(2)}$ primaries
is obtained by taking the polynomial
$P_N$ in \nsstates\ to be independent of $\psi^A$,
$j^A$ (and their derivatives). 
The resulting operators have spacetime scaling
dimension $h=j+1$ with $j$ determined by \dddii\ \gks.   
 
Similarly, one can study the transformation properties
of physical states under $\widehat{SU(2)}$, $\widehat{
U(1)}^4$ (see \gks\
for details). We would like to point out an interesting
property of the spectrum of the $\widehat{U(1)}^4$
symmetry in this model. 
{}From the form of the generators $\alpha^i_n$ \stvira\
it is clear that physical states which carry right-moving
momentum $\vec p$ along the $T^4$, such as \nsstates, and are
primary under affine $U(1)^4$, transform as:
\eqn\transuone{[\alpha^i_n, V_{\rm phys}^{\vec p}(j;m,\bar m)]=
p^iV_{\rm phys}^{\vec p}(j;m+n,\bar m)}
\ie\ the charges of primaries under $U(1)^4$ in spacetime
and on the worldsheet are the same. However, we saw in eq.
\virfull\ that the spacetime $U(1)^4$ generators are not 
normalized canonically. In terms of canonically normalized
generators, the charges of physical states are in fact
$p^i/\sqrt{p}$, where $p$ is given in \pdef. 
For example, if the worldsheet
$T^4$ on which the theory lives is a product of four circles
with radii $R_il_s$ and
the $B$ field on $T^4$ vanishes, the spectrum of $U(1)^4$
charges becomes 
\eqn\strspec{Q^i={1\over\sqrt{p}}\left({n\over R_i}+
{mR_i\over2}\right)}
It has been proposed that the spacetime SCFT corresponding to string 
theory on \adst\ is a blowing up deformation of the $(4,4)$ 
superconformal $\sigma$--model on the symmetric product $\MM^L/S_L$
with $L=kp$. For $\MM=T^4$ the symmetric product SCFT has a 
$\widehat{U(1)}^4_R\times\widehat{U(1)}^4_L$ affine symmetry, 
like string theory on \adst. The spectrum of charges under this
affine symmetry in the symmetric product SCFT is independent of 
the blowing up deformations. Therefore, it is interesting to compare 
it to the spectrum \strspec\ in string theory. As there,
we will for simplicity take the spacetime $T^4$ to
be a product of four circles with radii $r_i$ $(i=1,\cdots, 4)$ 
(which may in general be different from $R_i$), and set the $B$ 
field to zero. The generalization to an arbitrary $T^4$ is 
straightforward. 

To compare to \strspec\ we have to normalize the $\widehat{U(1)}^4$
generators in the symmetric product canonically, and then compute
their spectrum at the orbifold point. The resulting spectrum of 
charges $q_i$ is:
\eqn\symspec{q^i={1\over\sqrt{L}}\left({n\over r_i}+
{mr_i\over2}\right)}
Since $L=kp$ and both in \strspec\ and in \symspec\ 
states with all possible values of the integers $n,m$
appear, we conclude that
the string theory spectrum is incompatible
with the symmetric product at {\it any} point in its moduli
space (for $k>1$). 
We view this as evidence that the spacetime SCFT relevant
for string theory on \adst\ is not on the moduli space of
the symmetric product SCFT (at least for $\MM=T^4$ and $k>1$).

We next turn to a discussion of the chiral primaries. 
Unitarity of the spacetime $N=4$ superconformal algebra
implies that the scaling dimension 
$h$, and $SU(2)$ spin $j'$ satisfy the inequality
$h\ge j'$. Chiral primaries are states that
saturate this bound. One can prove
that the spectrum \nsstates\ satisfies
the $N=4$ unitarity bound. The proof uses
the worldsheet unitarity bounds $j<k/2, j'\leq k/2$. 
One also finds that the states that saturate the bound
are those with $N=1/2$ and $\vec p=0$ in \nsstates.
There are analogous statements in the Ramond sector. 

The chiral primaries can be analyzed
holomorphically; then the two chiralities
are combined 
in all possible ways consistent with
modular invariance. Consider first the NS sector. For $N=1/2$,
$\vec p=0$, \nsstates, \dddii\ imply that $j=j'$. For generic $j$
there are eight physical states (with given $m$, $m'$):
\eqn\masslessst{\eqalign{ 
\VV^i_j=&e^{-\phi}\lambda^iV_jV'_j\cr
\WW^\pm_j=&e^{-\phi}(\psi V_j)_{j\pm1}V'_j\cr
\XX^\pm_j=&e^{-\phi}V_j(\chi V'_j)_{j\pm 1}\cr
}}
where we have used results and notations
from Appendix A for combining fermions and bosons
into representations of the total current algebras,
and suppressed the indices $m,m',\cdots$.
One can show that the states \masslessst\ are
BRST invariant and thus physical while the two remaining
combinations, which involve $(\psi V_j)_j$ and
$(\chi V'_j)_j$ are not physical (one combination is
not BRST invariant, and the other is null).
The four operators $\VV_j$ have $h=j+1$, $j'=j$;
$\WW^\pm$ have $h=j+1\pm 1$, $j'=j$; and
$\XX^\pm$ have $h=j+1$, $j'=j\pm 1$. 
The chiral operators are thus $\WW_j^-$, $\XX_j^+$.  
The former are physical for $j\ge 1/2$ (for $j=0$
the corresponding operator is null); the latter
exist for all $j\ge 0$. 

One can similarly analyze the spectrum of chiral primaries
in the Ramond sector. It is convenient to split the spinors
into six dimensional spinors $S$,
which transform in the ${\bf(2,2)}$ of $SL(2)\times SU(2)$,
and spinors on the $T^4$, $\exp {i\over2}(\epsilon_4 H_4+\epsilon_5
H_5)$ (see Appendix A). Equating the spacetime scaling
dimension and $SU(2)$ spin leads to two
BRST invariant 
chiral primaries with $h=j'=j+1/2$, $j=0,1/2,\cdots$:
\eqn\rrpr{\YY^\pm_j=e^{-{\phi\over2}}e^{\pm{i\over2}(H_4-H_5)}
(SV_jV'_j)_{j-{1\over2}, j+{1\over2}}}
By applying the supercharges \supchh\ one finds that
the upper component of these superfields are 
$\VV^i_j$ \masslessst.

To summarize, the holomorphic analysis leads to the
following spectrum of chiral primaries:  
\bigskip
\vbox{
$$\vbox{\offinterlineskip
\hrule height 1.1pt
\halign{&\vrule width 1.1pt#
&\strut\quad#\hfil\quad&
\vrule width 1.1pt#
&\strut\quad#\hfil\quad&
\vrule width 1.1pt#
&\strut\quad#\hfil\quad&
\vrule width 1.1pt#\cr
height3pt
&\omit&
&\omit&
&\omit&
\cr
&\hfil ${\cal V}$&
&\hfil $h$&
&\hfil range of $j$&
\cr
height3pt
&\omit&
&\omit&
&\omit&
\cr
\noalign{\hrule height 1.1pt}
height3pt
&\omit&
&\omit&
&\omit&
\cr
&\hfil $\XX_j^+$&
&\hfil $j+1$&
&\hfil $0,1/2,\ldots$&
\cr
height3pt
&\omit&
&\omit&
&\omit&
\cr
\noalign{\hrule}
height3pt
&\omit&
&\omit&
&\omit&
\cr
&\hfil $\WW_j^-$&
&\hfil $j$&
&\hfil $1/2,1,\ldots$&
\cr
height3pt
&\omit&
&\omit&
&\omit&
\cr
\noalign{\hrule}
height3pt
&\omit&
&\omit&
&\omit&
\cr
&\hfil $\YY_j^\pm$&
&\hfil $j+1/2$&
&\hfil $0,1/2,\ldots$&
\cr
height3pt
&\omit&
&\omit&
&\omit&
\cr
}\hrule height 1.1pt
}
$$
}
\centerline{\sl Table 2: Holomorphic chiral primary operators.}
\def\Lallstates{Table 2}
Tensoring \Lallstates\ with its antiholomorphic
analog leads to the following list of chiral primaries: 
\bigskip
\vbox{
$$\vbox{\offinterlineskip
\hrule height 1.1pt
\halign{&\vrule width 1.1pt#
&\strut\quad#\hfil\quad&
\vrule width 1.1pt#
&\strut\quad#\hfil\quad&
\vrule width 1.1pt#\cr
height3pt
&\omit&
&\omit&
\cr
&\hfil $( h , {\bar h})$&
&\hfil degeneracy&
\cr
height3pt
&\omit&
&\omit&
\cr
\noalign{\hrule height 1.1pt}
height3pt
&\omit&
&\omit&
\cr
&\hfil $({l+3\over 2},{l+1\over 2})$&
&\hfil $1$&
\cr
height3pt
&\omit&
&\omit&
\cr
\noalign{\hrule}
height3pt
&\omit&
&\omit&
\cr
&\hfil $({l+1\over 2},{l+3\over 2})$&
&\hfil $1$&
\cr
\noalign{\hrule}
height3pt
&\omit&
&\omit&
\cr
&\hfil $({l+1\over 2},{l+1\over 2})$&
&\hfil $5$&
\cr
\noalign{\hrule}
height3pt
&\omit&
&\omit&
\cr
&\hfil $({l+2\over 2},{l+2\over 2})$&
&\hfil $1$&
\cr
\noalign{\hrule}
height3pt
&\omit&
&\omit&
\cr
&\hfil $({l+2\over 2},{l+1\over 2})$&
&\hfil $4$&
\cr
\noalign{\hrule}
height3pt
&\omit&
&\omit&
\cr
&\hfil $({l+1\over 2},{l+2\over 2})$&
&\hfil $4$&
\cr
height3pt
&\omit&
&\omit&
\cr
}\hrule height 1.1pt
}
$$
}
\centerline{\sl Table 3: Spectrum of chiral primaries for $AdS_3\times 
S^3\times T^4$.}
\def\fivedstates{Table 3}
\noindent where the index $l=0,1,\ldots $ up to an upper bound that
is determined by the fact that in \Lallstates, $j$ satisfies
the constraint $j\leq k/2-1$. 

The five $(h,\bar h)=({1\over2}, {1\over2})$ chiral primaries
in \fivedstates\ give rise to twenty moduli parametrizing the
space
\eqn\modnsf{SO(5,4;\ZZ)  \backslash SO(5,4)/ SO(5)\times SO(4)}
Sixteen of the moduli arise from the NS-NS sector on
the worldsheet; they correspond to the metric and $B$
field on the $T^4$. The remaining four are RR moduli.

\lref\malstr{J. Maldacena and A. Strominger, hep-th/9804085.}%
\lref\adsspe{S. Deger, A. Kaya, E. Sezgin, and P. Sundell, 
hep-th/9804166.}%

The spectrum of spacetime chiral primaries, 
\fivedstates, is in agreement 
with expectations from 
supergravity~\refs{\malstr,\adsspe,\flspectrum,\dbspectrum}. In fact,
\fivedstates\ is identical to a table that appears in 
\refs{\flspectrum}. The upper bound on $h, \bar h$
that arises from string theory can not be seen in the supergravity
analysis since it is due to string scale physics. 
Of course, the operators in \fivedstates\ are only 
those that correspond to the single
particle chiral states, and there are many additional
chiral operators that correspond to multiparticle states.
Finally, general considerations lead one to expect
the spectrum of chiral primaries to be truncated at $h,\bar h$ of 
order the central charge \malstr\ , but checking this 
directly may be beyond the reach 
of our perturbative analysis. 

\subsec{Heterotic strings on $AdS_3\times S^3\times T^4$}

The discussion of the previous subsections generalizes
easily to the heterotic case. The right-moving worldsheet
theory is the same as before, while the left-movers
are purely bosonic and live on the manifold
$AdS_3\times S^3\times T^4\times \Gamma^{16}$,
where $\Gamma^{16}$ is the $E_8\times E_8$
or ${\rm Spin}(32)/\ZZ_2$ torus\foot{Of course,
as usual, one can turn on moduli that mix the $T^4$
with $\Gamma^{16}$ and give rise to the standard Narain
moduli space of vacua.}. The symmetry structure
in the right-moving sector is as before, while
the left-movers give rise to non-supersymmetric conformal
symmetry, an $SU(2)$ affine Lie algebra from the sphere, 
and some additional affine symmetries from the $T^4$ and
$\Gamma^{16}$. At generic points in the Narain moduli space
there is a $U(1)^{20}\times SU(2)$ left-moving affine symmetry. 
At points with enhanced gauge symmetry, the affine symmetry
in spacetime is enhanced. 

The spacetime theory is thus a $(4,0)$ superconformal
field theory. As we saw \stvira\ -- \centchgs,
the spacetime central charge is determined by the {\it total}
$\widehat{SL(2)}$ level on the worldsheet, which is
$k=(k+2)-2$ for the right-movers, and $k+2$ for the left-movers.
Therefore, the central charges of the spacetime theory are:
$\bar c_{\rm st}=6pk$ and $c_{\rm st}=
6p(k+2)$. The difference
\eqn\diffc{c_{\rm st}-\bar c_{\rm st}=12p}
between the left and right central charges
is an example of a stringy correction to the
semiclassical results of \brhen, \cspacetime. In the supergravity
limit $c_{\rm st}=\bar c_{\rm st}$; \diffc\ is suppressed relative to 
the leading contribution by two powers of $l_s/l$. Similarly, the
levels of the left and right-moving $\widehat{SU(2)}$ algebras
in spacetime differ: $\bar k_{\rm st}-k_{\rm st}=2p$.

One can repeat the discussion of the previous subsections
rather closely; we will not describe the details here. 
As mentioned above,
the theory described in this subsection
is S-dual to type IIA on $AdS_3\times S^3\times K3$.
It might be interesting to study this duality 
further using the techniques of \gks\ and this paper.

\newsec{Conformal Field Theory on $S^3/\ZZ_N$}
\lref\gepnernotes{D. Gepner,
in proceedings, {\it Superstrings}, 238-302, Trieste, 1989.}%
As we saw in section 2, the
near-horizon geometry of $NS5$-branes and KK monopoles
includes a Lens space. In order to study string propagation
in this geometry we need to construct worldsheet CFT on
$SU(2)_R\times SU(2)_L/\ZZ_N$, where the $\ZZ_N$ orbifold
acts on the left-movers only. These CFT's were studied in 
\refs{\gps, \johnson}; in this section we will review their
properties, first in the bosonic and then in the supersymmetric 
case. 

\subsec{Bosonic CFT on Lens spaces}

Consider $SU(2)$ WZW theory at level $k$. 
The theory has two affine $SU(2)$ symmetries
acting on the left and right-movers. Denoting the
left-moving currents\foot{Whose OPE is given in Appendix A.}
by $K^a(z)$, and the right-moving ones by $\bar K^a(\bar z)$, the action
of the $\ZZ_N$ symmetry by which we mod out to get the Lens space
is as follows. We define a $2\times 2$ (hermitian) matrix of currents
$K\equiv K^a\sigma^a$, $\bar K\equiv \bar K^a\sigma^a$ and
$G\equiv\exp({2\pi i\over N}\sigma_3)$. The $\ZZ_N$ then acts on
the currents as:
\eqn\znaction{\eqalign{
&K\to GK G^{-1};\;\;\;\bar K\to \bar K\cr
&K^\pm\to e^{\pm{4\pi i\over N}}K^\pm, K^3\to K^3;\;\;\;
\bar K^a\to\bar K^a\cr
}}
Thus, for $N>2$ the projection breaks $SU(2)_L\to U(1)$.
One can think of the resulting asymmetric orbifold CFT in the
following way. It is well known (see \eg\ \cftbook)
that one can decompose $SU(2)$
WZW CFT into a product of the $SU(2)/U(1)$ parafermion coset CFT 
and a free scalar field $\xi$, 
normalized as $\langle\xi(z)\xi(0)\rangle=-{2\over k}\ln z$,
which is related to $K^3$ via:
\eqn\gaugecur{ K^3 = {ik\over 2}~\partial\xi}
Since the
eigenvalues of $K^3$ are half-integer,
$\xi$ is a compact scalar field, 
\eqn\compxi{\xi\sim \xi+4\pi}
More precisely, $\xi$ has period $2\pi$, but wavefunctions
with $j\in \ZZ+1/2$ go to minus themselves as $\xi\to\xi+2\pi$. 
Under the above decomposition, the $SU(2)$ primaries can be written 
as:
\eqn\para{V'_{j;m,\bar m}=f_{j;m,\bar m} e^{im\xi}}
$m$, $\bar m$ take values in the range
$|m|, |\bar m|\le j$,
with $j-m, j-\bar m\in \ZZ$ and $k\ge 2j\in \ZZ_+$. 
The chiral parafermion 
fields $f_{j;m, \bar m}$ commute with the current $K^3$ (but not with 
$\bar K^3$) and have scaling dimensions
\eqn\paradims{\eqalign{
\Delta=&{j(j+1)\over k+2}-{m^2\over k}\cr
\bar\Delta=&{j(j+1)\over k+2} \cr
}}
The left-moving 
parts of the operators $f_{j;m,\bar m}$ \para\ with $|m|=j$ 
are further identified, 
$(j,m)\simeq ({k\over2}-j, m\pm {k\over2})$.  
The currents $K^\pm$ take the form:
\eqn\currpara{
K^\pm=e^{\pm i\xi}\psi_{\rm para}^\pm}  
where $\psi_{\rm para}^\pm$ are parafermionic fields with scaling
dimension $(k-1)/k$. Comparing \currpara\ to \znaction\ 
we see that the $\ZZ_N$ acts on the chiral scalar field
$\xi$ as $\xi\to\xi+ 4\pi/N$, decreasing the radius
of $\xi$ \compxi\ by $N$ units. Out of the 
original operators in the $SU(2)$ WZW model \para\ only
those with $2m\in N\ZZ$ survive the $\ZZ_N$ projection.  
These form the untwisted sector of the orbifold.

The twisted sector of the orbifold includes operators 
of the form 
\eqn\twistop{V'_{j;m,m',\bar m}=
f_{j;m,\bar m}e^{im'\xi}}
with $m\not=m'$ (compare to \para). 
These must have the same periodicity
in $\xi$ as the untwisted ones: as $\xi\to\xi+2\pi$
they must be multiplied by $(-)^{2j}$. Therefore, 
they must satisfy:
\eqn\cccsss{m-m'\in \ZZ} 
Modular invariance requires the inclusion in the theory
of operators of the form \twistop\ that 
satisfy level matching and are mutually local
with respect to all operators in the untwisted sector \para\
that survive the $\ZZ_N$ projection. Requiring mutual locality
of \para\ (with $2m\in N\ZZ$) and \twistop\ leads to the
constraint
\eqn\newcons{m-m'\in {k\over N}\ZZ} 
Since completeness of the OPE requires us to include
{\it all} operators \twistop\ that satisfy \newcons, 
comparing to \cccsss\ we conclude that the asymmetric orbifold
we are studying can only be consistent when
${k\over N}\equiv N'\in \ZZ$. Level matching of the operators
\twistop\ further leads to the requirement,
\eqn\delth{{m^2\over k}-{m'^2\over k}\in \ZZ}
which together with \newcons\ means that $m$, $m'$
satisfy the constraints:
\eqn\intcond{\eqalign{ 
m'+m \in &~ N\ZZ \cr
m'- m \in &~ N'\ZZ \cr
}}
Clearly, the constraint \intcond\ must hold for all
operators in the orbifold theory, since we can generate
all such operators by multiplying operators of the form
\twistop. 
As a check on \intcond\ one can compute the resulting torus
partition sum and check its modular invariance. This is done
in Appendix B.
 
\subsec{$N=1$ SCFT on Lens spaces}
\lref\greenenotes{B. Greene, hep-th/9702155, in proceedings of TASI 1996, 
Boulder, CO.}%

The starting point of the construction is the $N=1$ superconformal 
$SU(2)$ WZW model (discussed around eq. \Jp). The total left-moving 
$SU(2)$ current $K^a$ has level $k$; the $\ZZ_N$ orbifold acts on
$K^a$ as before \znaction. To preserve the worldsheet  $N=1$ 
superconformal algebra \gsl\ the orbifold must also act on the 
left-moving fermions $\chi^a$ in a similar way to \znaction:
\eqn\actfer{\chi^\pm\to e^{\pm{4\pi i\over N}}\chi^\pm, \chi^3\to \chi^3}
and, as before, the right-movers are invariant.
To perform the orbifold  \znaction, \actfer,  
we decompose the $N=1$ superconformal
$SU(2)$ WZW model into $SU(2)/U(1)\times U(1)$. The first factor is well
known to have an accidental $N=2$ superconformal symmetry; it gives rise 
to an $N=2$ minimal model (see \eg\ \refs{\gepnernotes,\greenenotes}
for reviews). The second factor corresponds to a free superfield
$(\xi, \chi^3)$ where $\xi$ is related to $K^3$ by \gaugecur. 
The analogs of the spectrum generating operators \twistop\ in this case
are: 
\eqn\twistvert{ V'_{j;m,m',\bar m}=V_{j;m,\bar m}^{N=2}~e^{im'\xi}}
where $m$, $m'$ satisfy \intcond, and $V_{j;m,\bar m}^{N=2}$ 
are minimal model primaries on the left and $\widehat{SU(2)}$
primaries on the right. Focusing on the left-moving structure,
there are two kinds
of such operators: NS sector operators which are mutually
local with respect to the supercurrent \gsl, and Ramond
sector fields which have branch cuts with respect to $G$.
They have the spectrum of scaling dimensions $(\Delta)$
and $U(1)$ charges $(Q)$:
\eqn\specns{\eqalign{\Delta_{NS}=&{j(j+1)\over k}-
{m^2\over k};\;\;\;
Q_{NS}={2m\over k}\cr
\Delta_{R}=&{j(j+1)\over k}-{m^2\over k}+{1\over 8};\;\;\;
Q_{R}={2m\over k}\mp{1\over2}\cr
}}
where the range of the indices $j,m$ is as in \paradims\
in the NS sector, while in the Ramond sector $|m\mp{1\over2}|
\leq j$ (with the same identifications as before).
The two kinds of Ramond operators in \specns\ are related
by application of $G_0^\pm$, the zero modes of the $N=2$ 
superconformal generators. One can again verify using the
appropriate characters that the resulting theory is modular
invariant. This is discussed in Appendix B. 

\newsec{String Theory on $AdS_3\times S_3/\ZZ_N\times T^4$}

Armed with an understanding of string theory on $AdS_3$
(section 3), and CFT on $S^3/\ZZ_N$ (section 4) we are
now ready to tackle the problem of interest, string theory
on the manifold \backgg. Some interesting aspects of the
problem appear already in the bosonic case, which
we therefore discuss first. 

\subsec{Bosonic string theory on $AdS_3\times S^3/\ZZ_N\times
T^{20}$}

The right-movers are
described by CFT on $AdS^3\times S^3\times T^{20}$;
in the left-moving sector we replace the $S^3$ by the monopole
background described in section {\it 4.1}. To conform with
the supersymmetric case, we take the level of $\widehat{SL(2)}$
to be $k+2$ and that of $\widehat{SU(2)}$ to be $k-2$, so that
the total worldsheet central charge is $k$ independent.
The spacetime central charge \centchgs\ is in this case
$c_{\rm st}=\bar c_{\rm st}=6(k+2)p$. The torus
partition sum from which one can read-off the spectrum 
of physical states is (see \partsum\ for the notation):
\eqn\bosptsm{Z(\tau,\bar \tau)=
{1\over\sqrt{\tau_2}|\eta(\tau)|^2}
\sum_{j,\bar j\le {k\over2}-1}
\NN_{j,\bar j}\chi_j^{(k-2)}(\tau) 
\bar\chi_{\bar j}^{(k-2)}(\bar\tau)
{1\over |\eta(\tau)|^{40}}
\sum_{(\vec p,\vec{\bar p})\in 
\Gamma^{20,20}}q^{{1\over2} 
{\vec p}^2}{\bar q}^{{1\over2} {\vec 
{\bar p}}^2}} 
where the left-moving characters $\chi_j^{(k-2)}$ are
the bosonic monopole characters $\chi_j^{\rm monopole}$
described in Appendix B, while the right-moving ones
$\bar\chi_{\bar j}^{(k-2)}$ are $SU(2)$ WZW characters.
Of course, the bosonic theory is tachyonic, but it
is still useful for studying in a simpler setting
aspects of the theory
which survive in the supersymmetric examples.

One such aspect is the transformation of string
excitations under the $SO(3)$ rotation symmetry of the
two-sphere discussed in section 2. 
The sum over $j$ in the partition sum \bosptsm\
runs over both half-integer and integer
values; at the same time the theory only has bosonic
excitations -- all states contribute with positive
sign to the vacuum energy \bosptsm. 
Naively this implies that this background 
violates the spin-statistics theorem. 

\lref\coleman{S. Coleman, {\it The Magnetic Monopole Fifty
Years Later}, in proceedings, Int. School of subnuclear physics,
Erice 1981, Summer school in theoretical physics, Les Houches 1981, 
Summer Institute on Particle and Fields, Banff 1981.}%
To resolve this puzzle we need to recall some facts about the 
behavior of magnetic monopoles and dyons in quantum mechanics 
(see \coleman\ for a discussion). Consider a quantum mechanical 
particle with spin zero and electric charge 
$e$ in the background of a magnetic monopole with magnetic
charge $g$. Dirac quantization is the statement that 
$eg\in \ZZ/2$. The system has a conserved angular
momentum, but the allowed values that this angular momentum can
take are:
\eqn\angmom{j=eg, eg+1, \cdots}
Thus, the spin $j$ is bounded from below by $eg$ and
can even take half-integer values (if $eg\in 
\ZZ+1/2$). This is consistent with spin-statistics since,
as explained in \coleman, dyons with electric charge $e$ 
and magnetic charge $g$ with half integer $eg$ indeed behave
as fermions. In particular the wavefunction of the system
is antisymmetric under interchange of two such objects.  

This seems to explain half of the puzzle raised above.
The fact that the partition sum \bosptsm\ has contributions
with half-integer $j$ despite the fact that we expect all 
excitations of a bosonic string to have integer spin must be due to
the effect of the monopole field \angmom. In essence, 
some of the angular momentum $j$ is due to the
electromagnetic field of the dyon; it is thus
a property of the string vacuum and is
not intrinsic to a particular excitation with charge $e$.
In fact, we can use
this interpretation to compute the electric charge 
of the different string excitations. 
Recall \fixu\ that the magnetic charge of the vacuum
is $P_5$. The angular momentum $j$ of string states
is bounded by the quantum number $m$ \paradims, \twistop\ by
$j\ge m$. Comparing to \angmom\ we see that the electric
charge $e$ of states like \para, \twistop\ is
$e=m/P_5$. 

There still seems to be some tension between the 
fact that dyons with half integer $eg$ are fermions,
and the fact that excitations with half integer $j$
contribute with positive sign to the string partition 
sum \bosptsm. The resolution is that string theory on
$AdS_3\times S^3/\ZZ_N$ is not really quantizing these 
dyons. Instead it is quantizing bosonic particles 
{\it in the background} of a monopole with a given 
magnetic charge. Thus, the relevant statistics is not
that associated with exchange of dyons, but rather the one
that has to do with exchanging fluctuations of the bosonic
fields that correspond to string modes in the fixed background
of a monopole. The latter statistics is that of the 
basic quanta of the string field, which are bosonic.
Hence, there is no contradiction between \bosptsm\
and the spin-statistics theorem.  

\subsec{Heterotic string theory on $AdS_3\times S^3/\ZZ_N
\times T^4$}

We next turn to the heterotic theory. Our main task will
be to reproduce and extend the supergravity result,
\sugrastates, for the spectrum of chiral primaries. 

The basic structure is very similar to that of section
{\it 3.3}. The worldsheet right-movers are described
by a superconformal $\sigma$-model on 
$AdS_3\times S^3\times T^4$ and give rise to
$N=4$ superconformal symmetry in spacetime. 
The left-movers are described by a bosonic
worldsheet $\sigma$-model  
on $AdS_3\times S^3/\ZZ_N\times T^{20}$ and
give rise to conformal symmetry in spacetime. 
The lattice
of momenta corresponding to $T^4_R\times T^{20}_L$
is an even, self-dual Narain lattice $\Gamma^{4,20}$.
The left and right moving spacetime central charges are again
determined by the total levels of $\widehat{SL(2)}$
and are equal to: $c_{\rm st}=6p(k+2)$, $\bar c_{\rm st}=
6pk$. To determine the relation between $k$ and the number
of branes in the background, we recall that according
to section {\it 4.1} the level of the bosonic $\widehat{SU(2)}$,
$k-2$, must be a product of two integers 
\eqn\quantknn{k-2=NN'} 
in order for the asymmetric $\ZZ_N$ orbifold to be consistent.
We readily identify $N$ with the number of KK monopoles, $N'$
with the number of $NS5$-branes (see \cymetric\ -- \quant).
Thus, the central charge is in this case
\eqn\chet{\eqalign{
c_{\rm st}=&6p(NN'+4)\cr  
\bar c_{\rm st}=&6p(NN'+2)\cr  
}}
Comparing to the Brown and Henneaux semiclassical gravity
analysis \cssp\ we see that this is another example of stringy
corrections modifying the semiclassical formula \cspacetime. 

We next turn to the spectrum of chiral primaries under
the $(4,0)$ superconformal spacetime symmetry. The
spacetime supersymmetry in this model arises purely
from the right-movers. As in section 3, the most convenient
way to analyze the spectrum of chiral operators is to
construct it holomorphically and then combine 
the two worldsheet chiralities. 
Short representations of the right-moving
$N=4$ superconformal algebra are given in
\Lallstates. What remains to do is to analyze the left-moving
Virasoro primaries that can be tensored with \Lallstates\
to give physical states. We will next do that for the case
of the $A$-series modular invariant, $\NN_{j,\bar j}=\delta_{j,\bar j}$.
The generalization to other modular invariants is straightforward.
 
Since we want to compare to \sugrastates, we start
with chiral primaries arising from the untwisted sector
\para. The right-moving structure constrains $j=j'$ (the
bosonic $SL(2)$ and $SU(2)$ spins are the same).
One finds the following physical operators:
\eqn\hetchst{\eqalign{
&(J^A V_j)_{j\pm1} V'_{j;m},\;\;
2m\in N\ZZ;\;\;\;h=j+1\pm1\cr
&V_j(K^a V'_j)_{j\pm 1;m},\;\;
2m\in N\ZZ;\;\;\;h=j+1\cr
&B^\alpha V_jV'_{j;m},\;\;
2m\in N\ZZ,\;\;\alpha=1,\cdots, n_V-2;\;\;\;
h=j+1\cr
}}
The operators with $h=j$ in the first line of \hetchst\ start
only at $j=1/2$. The same is true for those with $SU(2)$ spin
$j-1$ in the second line. The rest of the towers of operators start
at $j=0$. The notation in the last line of \hetchst\ is as follows.
$B^\alpha$ are purely left-moving worldsheet currents. At generic points
in the Narain moduli space of $\Gamma^{4,20}$ there are twenty
such currents, corresponding to $\partial X^i$, with $i$ running
over the directions of the $T^4$, and the sixteen Cartan subalgebra
generators of the heterotic gauge group in ten dimensions. Thus,
generically, $n_V=22$. At points in the Narain moduli space where
the gauge symmetry is enhanced to a non-abelian group $G$,
$n_V$ increases
accordingly. At such points 
the spacetime theory has a current algebra $\hat G$ 
at level $p$.

The holomorphic operators \hetchst\ 
can be arranged into the following towers: 
\bigskip
\vbox{
$$\vbox{\offinterlineskip
\hrule height 1.1pt
\halign{&\vrule width 1.1pt#
&\strut\quad#\hfil\quad&
\vrule width 1.1pt#
&\strut\quad#\hfil\quad&
\vrule width 1.1pt#
&\strut\quad#\hfil\quad&
\vrule width 1.1pt#\cr
height3pt
&\omit&
&\omit&
&\omit&
\cr
&\hfil $h$&
&\hfil degeneracy&
&\hfil range of $j$&
\cr
height3pt
&\omit&
&\omit&
&\omit&
\cr
\noalign{\hrule height 1.1pt}
height3pt
&\omit&
&\omit&
&\omit&
\cr
&\hfil $j+2$&
&\hfil $1$&
&\hfil $0,1,\ldots$&
\cr
height3pt
&\omit&
&\omit&
&\omit&
\cr
\noalign{\hrule}
height3pt
&\omit&
&\omit&
&\omit&
\cr
&\hfil $j$&
&\hfil $1$&
&\hfil $1,2,\ldots$&
\cr
height3pt
&\omit&
&\omit&
&\omit&
\cr
\noalign{\hrule}
height3pt
&\omit&
&\omit&
&\omit&
\cr
&\hfil $j+1$&
&\hfil $n_V-1$&
&\hfil $0,1,\ldots$&
\cr
height3pt
&\omit&
&\omit&
&\omit&
\cr
\noalign{\hrule}
height3pt
&\omit&
&\omit&
&\omit&
\cr
&\hfil $j+1$&
&\hfil $1$&
&\hfil $1,2,\ldots$&
\cr
height3pt
&\omit&
&\omit&
&\omit&
\cr
}\hrule height 1.1pt
}
$$
}
\centerline{\sl Table 4: Operators of bosonic sector of heterotic string. 
}
\def\Rallstates{Table 4}
The spectrum of chiral primaries of the heterotic
string is now obtained by tensoring \Rallstates\ with
\Lallstates. Furthermore, to compare to supergravity we
put the index $m$ in \Rallstates\ to zero, since the states
with $m\not=0$ are not low energy states in the 
supergravity limit $N,N'\to\infty$. 
This implies that the index $j$ (which is the same for left and 
right-movers) is integer. It is not difficult to see that the resulting
spectrum is 
precisely of the form indicated in \sugrastates\ with 
$n_H=2(n_V-1)$ 
and $n_S=2$. 

Moduli of the spacetime CFT arise from chiral primaries with
$\bar h=1/2$, $h=1$. In \sugrastates\ they arise from the
$n_H=42$ hypermultiplets that exist generically in moduli
space; these give rise to an $84$ dimensional
moduli space. In the string description the spacetime moduli
are directly related to the worldsheet moduli 
$\partial x^a\bar\partial \bar x^{\bar b}$, where 
$a$ runs over the sixteen chiral scalars which exist 
in the heterotic string already in ten dimensions, the
four scalars parametrizing the $T^4$, and $\xi$ defined 
in \gaugecur. The index $\bar b$ runs over the four
antiholomorphic scalars parametrizing the right-movers
on $T^4$. From the string point of view it is obvious
that the moduli space of spacetime SCFT's is
\eqn\modhetob{
SO(4,21;\ZZ)  \backslash SO(4,21)/SO(4)\times SO(21) 
}
since this is the moduli space of worldsheet theories.
Note that the moduli space \modhetob\ is consistent 
with the fact that, as explained in section 2, 
the radius of $x^4$, $R_4$ \fixx, 
is a fixed scalar in the near-horizon geometry \nearh.

While the string theory analysis reproduces the supergravity
result of \sugrastates\ in the region where the latter is
expected to be applicable, it also generalizes 
it significantly. We have already seen an example of this above:
the states visible in supergravity have $m=m'=0$ in the notation
of \twistop\ (and thus integer $j$) while the full string
spectrum has in addition states with $m=m'\in {N\over2}\ZZ$,
some of which have half-integer $j$ if $N$ is odd. Since T-duality
in $x^4$ exchanges $N$ and $N'$ (see section 2), we also 
expect to find similar states in the twisted sector with
$m=-m'\in {N'\over2}\ZZ$. 

\lref\dabhar{A. Dabholkar and J. Harvey, \prl{63}{1989}{478};
A. Dabholkar, G. Gibbons, J. Harvey and F. Ruiz Ruiz,
\np{340}{1990}{33}.}%
In fact, the twisted sectors have in this case
a rich spectrum of stringy chiral operators.
Consider, for example, vertex operators of the general form: 
\eqn\newchiral{(\bar\psi V_j)_{j-1} V'_{j;m,m',\bar m} 
P_n(\psi^A, J^A,\cdots)}
In the right-moving sector this is an operator of the type
$\WW_j^-$, as in \Lallstates; hence it is chiral under the
right-moving spacetime $N=4$ superconformal algebra.
$V'$ is given by \twistop. The left-moving
worldsheet scaling dimension $n$ of the polynomial $P_n$
is determined by level matching:
\eqn\levmatn{n-1={m^2-m'^2\over NN'}}
Recall that the r.h.s. of \levmatn\ is integer \delth,
since $m$, $m'$ must satisfy \intcond. 
For every solution of the level matching condition 
\levmatn\ one finds an operator in a short multiplet
of the spacetime $(4,0)$ superconformal symmetry. 
For large $n$ the growth of the number of such
solutions is exponential (in $\sqrt n$). 
These states are curved spacetime analogs of Dabholkar-Harvey
states \dabhar. Clearly, there are other states of
the general form \newchiral; their construction is a straightforward
application of the methods of this paper and we will not discuss
them in detail. 
 
The string construction also allows one to study in a simple way
the transformation properties of various operators under the
full Virasoro algebra. The supergravity analysis that leads to
\sugrastates\ is only sensitive to the transformation properties
of the different states under the global superconformal
algebra. While on the supersymmetric side this is not a 
serious restriction -- an operator with $h=j$ must be a
primary under the full superconformal algebra since (at least
in a unitary SCFT) it cannot be a descendant -- on the left-moving
bosonic side one could ask whether the states in \sugrastates\ are
primary under the full Virasoro algebra or only under its $SL(2)$
subgroup; the latter are known as quasi-primaries in $2d$ CFT.
 
This question can be answered in string theory by applying
the Virasoro generators \stvira\ to the vertex operators 
\hetchst\ and asking whether \primr\ is satisfied for all
$n$ or only for $n=0,\pm1$. One finds that all operators
are primary under the full Virasoro algebra except for those
on the first line of \hetchst\ that involve $(J^AV_j)_{j+1}$; 
these are quasi-primary. This is not surprising given the
fact that for $j=0$ this combination appears in the Virasoro
generator itself, and the stress tensor is a quasi-primary
descendant of the identity in $2d$ CFT. 

\subsec{Type II string theory on $AdS_3\times S^3/\ZZ_N\times T^4$}

The main difference with respect to the analysis of the previous 
subsection is that the left-movers now live on 
$AdS_3\times S^3/\ZZ_N\times T^4$ and are described
by a superconformal worldsheet theory. 
The spacetime central charge is in this case given 
exactly by the semiclassical results \cssp, 
$c_{\rm st}=\bar c_{\rm st}=6pk=6pNN'$.

In order to define the worldsheet theory we have to
specify the GSO projection on the worldsheet. Performing
a chiral projection, which acts separately on left and
right-movers, $(-)^{F_L}=(-)^{F_R}=1$, leads to spacetime
supersymmetry, which is still coming only from the right-moving
sector (in agreement with the brane picture of section 2).
The fact that the left-moving supercharges \supchh\ are
projected out can be seen by noting that
their values of $m, m'=\pm {1\over2}$ are inconsistent
with the projection \intcond\ when $N, N'>1$. To see the
problem in more detail, trying to enforce mutual locality
of \supchh\ with the twisted NS sector states \twistvert\
leads to the condition $(m-m')/k\in \ZZ$ or, using \intcond,
$1/N\in \ZZ$, which implies $N=1$ ($N'=1$ also has enhanced
SUSY, arising from the twisted sector). 

It is not difficult to repeat the analysis of chiral primaries
performed in the previous subsection for the type II case.
We start again with states that survive in the supergravity limit.
Their holomorphic, left-moving structure
is closely related to \rrpr, \hetchst. In the NS sector
one has (in the $-1$ picture):
\eqn\IIchst{\eqalign{
&e^{-\phi}(\psi^A V_j)_{j\pm1} V'_{j;m},\;\;
2m\in N\ZZ;\;\;\;h=j+1\pm1\cr
&e^{-\phi}V_j(\chi^a V'_j)_{j\pm 1;m},\;\;
2m\in N\ZZ;\;\;\;h=j+1\cr
&e^{-\phi}\lambda^iV_jV'_{j;m},\;\;
2m\in N\ZZ,\;\;i=1,\cdots,4;\;\;\;
h=j+1\cr
}}
where $V'$ is given by \twistvert\ with $m=m'$,
and we are temporarily suppressing the right-moving
index $\bar m$ (since the analysis is chiral).  
As in the previous subsection, one of the towers
on the first line of \IIchst\ and one of the 
towers on the second line (the ones corresponding
to spin $j-1$) start at $j={1\over2}$. The
other towers exist for all $j\geq 0$ (bounded
from above by the worldsheet unitarity bound). 
The Ramond sector states are (in the $-1/2$ picture):
\eqn\ramchi{e^{-{\phi\over2}}e^{\pm{i\over2}(H_4-H_5)}
(SV_jV'_j)_{j\pm{1\over2}, j\pm{1\over2};m},\;\;2m\in N\ZZ}
Altogether, the quantum numbers and degeneracies
of the left-moving parts of the chiral operators
which survive in the supergravity $(m=0)$ limit are: 
\bigskip
\vbox{
$$\vbox{\offinterlineskip
\hrule height 1.1pt
\halign{&\vrule width 1.1pt#
&\strut\quad#\hfil\quad&
\vrule width 1.1pt#
&\strut\quad#\hfil\quad&
\vrule width 1.1pt#
&\strut\quad#\hfil\quad&
\vrule width 1.1pt#\cr
height3pt
&\omit&
&\omit&
&\omit&
\cr
&\hfil $h$&
&\hfil degeneracy&
&\hfil range of $j$&
\cr
height3pt
&\omit&
&\omit&
&\omit&
\cr
\noalign{\hrule height 1.1pt}
height3pt
&\omit&
&\omit&
&\omit&
\cr
&\hfil $j+2$&
&\hfil $1$&
&\hfil $0,1,\ldots$&
\cr
height3pt
&\omit&
&\omit&
&\omit&
\cr
\noalign{\hrule}
height3pt
&\omit&
&\omit&
&\omit&
\cr
&\hfil $j$&
&\hfil $1$&
&\hfil $1,2,\ldots$&
\cr
\noalign{\hrule}
height3pt
&\omit&
&\omit&
&\omit&
\cr
&\hfil $j+1$&
&\hfil $5$&
&\hfil $0,1,\ldots$&
\cr
\noalign{\hrule}
height3pt
&\omit&
&\omit&
&\omit&
\cr
&\hfil $j+1$&
&\hfil $1$  &
&\hfil $1,2,\ldots$&
\cr
\noalign{\hrule}
height3pt
&\omit&
&\omit&
&\omit&
\cr
&\hfil $j+1/2$&
&\hfil $4$&
&\hfil $1/2,3/2,\ldots$&
\cr
\noalign{\hrule}
height3pt
&\omit&
&\omit&
&\omit&
\cr
&\hfil $j+3/2$&
&\hfil $4$&
&\hfil $1/2,3/2,\ldots$&
\cr
height3pt
&\omit&
&\omit&
&\omit&
\cr
}\hrule height 1.1pt
}
$$
}
\centerline{\sl Table 5: Left-moving primaries for $AdS_3\times 
S^3/\ZZ_N \times T^4$.}
\def\Rstates{Table 5}
The physical spectrum of chiral primaries is obtained by tensoring 
\Rstates\ with \Lallstates. The result is precisely
the one given in \sugrastates\ with $n_V=n_H=14$ and 
$n_S=6$, in agreement with supergravity.

In section {\it 2.5} we saw that the moduli space of vacua
of string theory on $AdS_3\times S^3/\ZZ_N\times T^4$ is
the twenty eight dimensional coset \modspc. It is interesting
to ask how this space arises in the worldsheet construction.
The twenty eight moduli are the upper components of chiral
primaries with $h=1$, $\bar h={1\over2}$. Twenty of them arise
from the NS sector. These are the obvious moduli of the
worldsheet theory; they parametrize the coset \modnsf. 
The remaining eight moduli come from the RR sector,
and enlarge the moduli space from \modnsf\ to \modspc.
As a check, note that $SO(5,4)$ is a subgroup of $F_{4(4)}$
and $SO(5)\times SO(4)$ is a subgroup of $USp(2)\times
USp(6)$. 

The situation is in fact very similar to that in type II
string theory on $T^5\times \IR^{4,1}$, 
where the full moduli space is given
by the coset \modesix\ but only the subspace 
$SO(5,5;\ZZ)\backslash SO(5,5)/SO(5)\times SO(5)$
comes from NS fields (the Narain moduli space) while the 
rest of the moduli come from the RR sector. 
$SO(5,5)$ is a subgroup of $E_{6(6)}$ and $SO(5)\times SO(5)$
a subgroup of $USp(8)$.

\noindent
Comments:
\item{(a)} As in the previous subsection, the string
analysis produces a rich spectrum
of stringy chiral operators in addition to those seen
in supergravity.
\item{(b)} Both here and in the two previous subsections
we have implicitly assumed that $N, N'>2$. When either
$N$ or $N'$ is two, the theory in fact has more symmetry,
both on the worldsheet and in spacetime. 
When either $N$ or $N'$ is one, we get back the theory
discussed in \gks.

\bigskip
\noindent{\bf Acknowledgements:}
We thank J. Harvey, C. Johnson, E. Martinec, N. Seiberg and F. Wilczek 
for discussions. The work of DK and FL is supported in part 
by DOE grant \#DE-FG02-90ER40560. 
FL is also supported by a McCormick fellowship through the EFI.
RGL is supported by a DOE Outstanding Junior Investigator Award, under
grant \#DE-FG02-91ER40677.

\appendix{A}{Some Properties of String Theory on $AdS_3\times S^3\times
T^4$}

In this Appendix we collect some conventions and results about
CFT and string theory on $AdS_3\times S^3\times T^4$. 

\noindent {\bf 1. Worldsheet fermions:}
The fermions
$\psi^A$, $\chi^a$, $\lambda^j$ defined in the text are normalized as:
\eqn\normferm{\eqalign{
\langle\psi^A(z)\psi^B(w)\rangle=&{k\eta^{AB}/2\over z-w},
\qquad A,B=+,-,3\cr
\langle\chi^a(z)\chi^b(w)\rangle=&{k\delta^{ab}/2\over z-w},
\qquad a,b=+,-,3\cr
\langle\lambda^i(z)\lambda^j(w)\rangle=&{\delta^{ij}\over z-w},
\qquad i,j=1,2,3,4\cr
}}
The metrics are $\eta^{33}=-1, \eta^{+-}=2$;
$\delta^{33}=1, \delta^{+-}=2$.

To study the Ramond sector of the worldsheet theory 
one needs to construct the spin fields for
$\psi^A$, $\chi^a$, $\lambda^j$. 
It is convenient to
choose a complex structure by pairing the
ten free fermions into five complex ones, and bosonizing
them into canonically normalized scalar fields, $H_I$,
(normal ordering implied):
\eqn{\LHdefs}{\eqalign{
-i\partial H_1 =& {1\over k}\psi^+\psi^-\cr
-i\partial H_2 =& {1\over k}\chi^+\chi^-\cr
i\partial  H_3 =& {2\over k}\psi^3\chi^3\cr
\partial  H_4=& \lambda^1\lambda^2\cr
\partial H_5=& \lambda^3\lambda^4
}}
The spin fields take the form $S=\exp{i\over2}\sum_{I=1}^5
\epsilon_IH_I$, where $\epsilon_I=\pm1$.

\noindent
{\bf 2. Worldsheet current algebra:}
The currents \Jp\ satisfy the OPE's:
\eqn\two{\eqalign{
J^A(z)J^B(w)=& {k\eta^{AB}/2
\over(z-w)^2}+{i\epsilon^{AB}_{\,\,\,\,\,\,C}J^C\over z-w}+\cdots,
\qquad A,B,C=+,-,3\cr
K^a(z)K^b(w)=&          
{k\delta^{ab}/2
\over(z-w)^2}+{i\epsilon^{ab}_{\,\,\,\,\,\,c}K^c\over z-w}+\cdots;
\qquad a,b,c=+,-,3\cr
}}
where $i\epsilon^{+-3} = 2 $ in both $SL(2)$
and $SU(2)$ (thus, lowering a $3$ index gives
a relative minus sign between the two). 

As in the text, we denote primaries of the bosonic 
$\widehat{SL(2)}$ by $V_{j;m,\bar m}$, and those of
the bosonic $\widehat{SU(2)}$ by $V'_{j;m,\bar m}$.
In both cases $j$ is the quadratic Casimir, while
$(m, \bar m)$ are the eigenvalues of $(j^3, \bar j^3)$ or 
$(k^3, \bar k^3)$, as appropriate. The normalizations of 
the operators $V$, $V'$ are such that:
\eqn\someopes{\eqalign{
j^\pm(z) V_{j;m,\bar m}(w)=&(m\mp j) V_{j; m\pm1,\bar 
m}(w)/(z-w)+\ldots\cr 
j^3(z) V_{j;m,\bar m}(w)=&m V_{j;m,\bar m}(w)/(z-w)+\ldots\cr 
k^\pm(z) V^\prime_{j;m,\bar m}(w)=&(j\mp m) 
V_{j; m \pm1,\bar m}(w)/(z-w)+\ldots\cr
k^3(z) V^\prime_{j;m,\bar m}(w)=&m 
V^\prime_{j;m,\bar m}(w)/(z-w)+\ldots \cr
}}
The scaling dimensions corresponding to $SL(2)_{k+2}$ and
$SU(2)_{k-2}$ are:
\eqn\scdim{\Delta(V_j)=-{j(j+1)\over k};\;\;\;
\Delta(V'_{j'})={j'(j'+1)\over k}}
The operators $V_j$ are left-right symmetric; thus their
left and right scaling dimensions are equal. For $SU(2)$,
there are non-diagonal modular invariants with $j'\not=\bar j'$.

The fermions $\psi^A$, $\chi^a$ transform in the spin 
$(1,0)$ and $(0,1)$ representations of
$SL(2)\times SU(2)$. It is useful to decompose the products
of fermions with bosonic primaries $V$, $V'$ into representations
of the total currents \Jp. The relevant decompositions are:
\eqn\sltwocg{\eqalign{
 (\psi V_j)_{j+1;m,\bar m} =& ~(j+1-m)(j+1+m)\psi^3 
V_{j;m,\bar m}+
{1\over 2}(j+m)(j+1+m)\psi^+ V_{j;m-1,\bar m} \cr +&~ 
{1\over 2}(j-m)(j+1-m)\psi^- V_{j;m+1,\bar m} \cr
(\psi V_j)_{j;m,\bar m} =& ~m\psi^3 V_{j;m,\bar m}-
{1\over 2}(j+m)\psi^+ V_{j;m-1,\bar m} 
+ {1\over 2}(j-m)\psi^- V_{j;m+1,\bar m} \cr
(\psi V_j)_{j-1;m,\bar m} =& ~\psi^3 V_{j;m,\bar m}-{1\over 2}\psi^+
V_{j;m-1,\bar m} - {1\over 2}\psi^- V_{j;m+1,\bar m}\cr 
(\chi V^\prime_j)_{j+1;m,\bar m} 
=&~ (j+1-m)(j+m+1)\chi^3 V^\prime_{j;m,\bar m}-
{1\over 2}(j+m)(j+m+1)\chi^+ V^\prime_{j;m-1,\bar m} \cr
+& {1\over 2}(j-m)(j+1-m)\chi^- V^\prime_{j;m+1,\bar m} \cr
(\chi V^\prime_j)_{j;m,\bar m} =& ~m\chi^3 V^\prime_{j;m,\bar m}+
{1\over 2}(j+m)\chi^+ V^\prime_{j;m-1,\bar m} + 
{1\over 2}(j-m)\chi^- V^\prime_{j;m+1,\bar m} \cr
(\chi V^\prime_j)_{j-1;m\bar m} 
=& ~\chi^3 V^\prime_{j;m,\bar m}+{1\over 2}\chi^+
V^\prime_{j;m-1,\bar m} -{1\over 2}\chi^- V^\prime_{j;m+1,\bar m} 
}}
Similarly, the eight spin fields 
$S=\exp{i\over2}(\epsilon_1 H_1 +\epsilon_2 H_2 +\epsilon_3 H_3 )$ 
associated with the fermions 
$\psi^A$, $\chi^a$ transform as two copies 
(corresponding to $\epsilon_1\epsilon_2\epsilon_3=\pm 1$) 
of $(1/2,1/2)$ under $SL(2)\times SU(2)$. 
The product $SV_jV'_{j'}$ can be decomposed into representations
of the total currents \Jp. This gives rise to the four
representations $(j\pm1/2, j'\pm1/2)$ for each of the two chiralities
of $S$ above. 

\noindent
{\bf 3. Wakimoto representation of $SL(2)$ CFT:}
A convenient representation of CFT on $AdS_3$ is given
by the Wakimoto representation \wakim. The worldsheet
Lagrangian is: 
\eqn\acwak{\LL=\partial\phi\bar\partial\phi-
{2\over\alpha_+}\hat R^{(2)}\phi+\beta\bar\partial
\gamma+\bar\beta\partial\bar\gamma-\beta\bar\beta\exp\left(
-{2\over\alpha_+}\phi\right)}
where\foot{Recall that the level of the bosonic
$\widehat{SL(2)}$ algebra is $k+2$.}
$\alpha_+^2=2k$ is related to $l$, the radius of
curvature of $AdS_3$, via:
\eqn\tb{l^2=l_s^2 k.}
Integrating out $\beta, \bar\beta$ leads to the standard
description of CFT on $AdS_3$. The description with $\beta,
\bar\beta$ \acwak\ is convenient since it simplifies in the limit 
$\phi\to\infty$: 
\item{(a)} The interaction term $\beta\bar\beta
\exp\left(-{2\phi\over\alpha_+}\right)$ goes to zero; thus
the worldsheet theory becomes weakly coupled.  
\item{(b)} The linear dilaton term implies that
the string coupling goes to zero exponentially in
$\phi$; thus the spacetime theory becomes weakly coupled.

\noindent
In the free field region $\phi\to\infty$
the propagators that follow from \acwak\ are:
$\langle\phi(z)\phi(0)\rangle=-\log|z|^2$,
$\langle\beta(z)\gamma(0)\rangle=1/z$.  The current algebra is
represented by (normal ordering is implied):
\eqn\curalg{\eqalign{
j^3=&\beta\gamma+{\alpha_+\over2}\partial\phi\cr
j^+=&\beta\gamma^2+\alpha_+\gamma\partial\phi
+k\partial\gamma\cr
j^-=&\beta.\cr}}
Bosonic primary vertex operators are given by:
\eqn\primr{V_{jm\bar m }=\gamma^{j+m}\bar \gamma^{j+ \bar m}
\exp\left( {2j\over\alpha_+}\phi\right).}
States \primr\ with $j>-1/2$ correspond
to wavefunctions on $AdS_3$ that are exponentially supported at
$\phi\to\infty$. Thus many of their properties, such as the
transformation properties under the infinite symmetry algebras
\stvira, \virfull\ can be studied using the above weakly coupled
representation.

\appendix{B}{Modular properties of monopole characters}

In this Appendix we discuss the modular transformation properties
of the characters of the monopole CFT discussed in section 4.
We start with the bosonic theory and then turn to the fermionic one.

\noindent {\bf 1. Bosonic case:}
The $SU(2)_k$ character
in the representation with spin $j$, $\chi_j^{(k)}(\tau)$,
is decomposed in terms of parafermion characters
$\Omega_{jm}^{(k)}(\tau)$ and $U(1)$ characters\foot{More
precisely $L_m^{(k)}={1\over\eta(q)}\sum_{n\in\ZZs}                     
q^{k(n+m/k)^2}$ 
is the character of an extended 
algebra that exists in $c=1$ CFT for rational $R^2/2$.
See \cftbook\ for a more detailed discussion.} 
$L_m^{(k)}(\tau)$ as:
\eqn\paradef{\chi^{(k)}_j(\tau) ={1\over2} \sum_{2m=-k+1}^k
\Omega_{jm}^{(k)}(\tau)L_m^{(k)}(\tau)
}
where it is understood that $j-m\in \ZZ$, and
we have extended the range of $m$ for convenience
to $m=-k+1,\cdots, k$. By using the identification
\eqn\identii{
\Omega^{(k)}_{j,m}=\Omega^{(k)}_{{k\over2}-j, m\pm {k\over2}}
}
one can always map $(j,m)$ to the fundamental range
discussed around eq. \paradims. Thus each distinct state
contributes twice to the r.h.s. of \paradef; hence the $1/2$. 
The monopole characters obtained after performing
the chiral $\ZZ_N$ twist \znaction\ can be written 
as 
\eqn\chimonbos{\chi^{\rm monopole}_j (\tau)={1\over2} 
\sum_{m'}\sum_{2m=-k+1}^{k}
\Omega^{(k)}_{jm}(\tau) L_{m'}^{(k)}(\tau)}
where \intcond\ $m'+m\in N\ZZ$, $m'-m\in N'\ZZ$.

We wish to determine the modular properties of the monopole 
characters \chimonbos. Since we have imposed level matching,
the transformation under $\tau\to\tau +1$
is the same as that of the corresponding $SU(2)_k$ character.
We next show that the same is true for the transformation 
$\tau\to -{1/\tau}$.

The $SU(2)_k$ characters satisfy~\refs{\cftbook}
\eqn\modree{\chi^{(k)}_{j}(-{1\over\tau}) = \sqrt{2\over k+2} 
\sum_{j^\prime=0}^{k\over2}
\sin\left[ {\pi (2j+1)(2j^\prime +1)\over k+2} \right] 
\chi^{(k)}_{j^\prime}(\tau)}
while for the $U(1)$ characters:
\eqn\modtheta{L_m^{(k)}(-{1\over\tau})={1\over\sqrt{2k}}
\sum_{2l=-k+1}^{k} e^{-{4\pi iml\over k}}{L_l^{(k)}(\tau)}}
These transformation properties, and the definition of the parafermionic
characters \paradef\  give
\eqn\modpara{\Omega^{(k)}_{jm}(-{1\over\tau}) = 
{1\over\sqrt{k(k+2)}} \sum_{j'=0}^{k\over2} \sum_{
m'=-k+1}^{k}
e^{4\pi i mm'\over k}\sin
\left[ {\pi (2j+1)(2j^\prime +1)\over k+2} 
\right]\Omega^{(k)}_{j'm'}(\tau)}
With these formulae in hand it is straighforward to compute the 
transformation properties of the monopole characters from their 
definition \chimonbos. After performing the various sums we find
\eqn\modmon{\chi_{j}^{\rm monopole}
(-{1\over\tau}) = \sqrt{2\over k+2} 
\sum_{2j^\prime=0}^{k}
\sin\left[ {\pi (2j+1)(2j^\prime +1)\over k+2} \right] 
\chi^{\rm monopole}_{j^\prime }(\tau)}
Comparing to \modree\ we see that the transformation
properties of the monopole characters at a given level are identical
to those of the $SU(2)$ characters at the same level. Therefore, we can 
take any modular invariant theory with $SU(2)_R\times SU(2)_L$ affine 
symmetry, and replace the $SU(2)$ characters by monopole characters for 
one of the two worldsheet chiralities without spoiling modular 
invariance. 

\noindent {\bf 2. Supersymmetric case:}
In the supersymmetric case, characters depend on
the spin structure
and $(-)^F$ projection
for the fermions. They can be distinguished\foot{In
the literature on $N=2$ models (see \eg\
\refs{\gepnernotes,\greenenotes}) 
sectors labeled by $s$ have insertions of $1\pm(-)^{F}$; they
are obvious linear combinations of ours.}
by an index $s=0,1,2,3$; 
$s=0,2$ correspond to the Ramond
sector with an insertion of $(-)^{sF/2}$. $s=1,3$
correspond to the NS sector with an insertion of $(-)^{(s-1)F/2}$
The monopole characters are
\eqn\chimonsusy{\chi^{\rm monopole}_{js} (\tau)= 
{1\over 2}\sum_{m,m'} \chi^{N=2}_{jms}(\tau) L_{m'}^{(k)}(\tau)
\left({\Theta_s\over\eta}\right)^{1\over2}}
with the sum over $m$, $m'$ 
running over the same range (and with the
same identifications) as in \chimonbos. $\chi^{N=2}_{jms}$
is the $N=2$ minimal model character with the appropriate
boundary conditions. $\Theta_s(\tau)$ are related to the standard
Jacobi $\theta$-functions by relabeling of indices: $s=0,2$
correspond to $\theta_2$, $\theta_1$, while $s=1,3$ correspond
to $\theta_3$, $\theta_4$. 
The last two terms on the r.h.s. of
\chimonsusy\ are the contributions of $K^3$ and $\chi^3$. 
Again, level matching implies that
\chimonsusy\ transforms under 
$\tau\to\tau + 1$ in the same way as the 
untwisted theory. 
The transformation under $\tau\to -{1/\tau}$ is 
discussed next.
 
In the untwisted $SU(2)$ WZW theory 
the characters are
\eqn\freechar{\chi^{SU(2)}_{js}(\tau) = \chi^{(k-2)}_j(\tau)
\chi^{(2)}_s(\tau)}
where $\chi^{(k-2)}_j$ are the $SU(2)_{k-2}$ characters and
\eqn\chiferm{\chi^{(2)}_s (\tau) =
 \left({\Theta_s\over\eta}\right)^{3/2}}
Using \modree\ and the transformation of the $\Theta$-functions 
\eqn\modthetaaa{{\Theta_s\over\eta}(-{1\over\tau})=\sum_{s'}\CC_{ss'}
{\Theta_{s'}\over\eta}(\tau),}
where the matrix elements of the symmetric matrix $\CC$
are $\CC_{ss'}=1$ for $(s,s')=(0,3)$, 
$(2,2)$, $(1,1)$ and $\CC_{ss'}=0$ otherwise, 
one finds for the untwisted theory
\eqn\chisusmod{\chi^{\rm SU(2)}_{js} (-{1\over\tau})
= {1\over\sqrt{2k}}\sum_{j^\prime s^\prime}\CC_{ss'}
\sin\left[ {\pi (2j+1)(2j^\prime +1)\over k} \right] 
\chi^{\rm SU(2)}_{j's'}(\tau).}

Returning to the monopole characters \chimonsusy,
the $N=2$ minimal model characters transform as
\eqn\modree{\chi_{jms}^{N=2}(-{1\over\tau}) = {1\over\sqrt{2k}} 
\sum_{j^\prime m' s^\prime}\CC_{ss'}
e^{{4\pi imm^\prime\over k}}
\sin\left[ {\pi (2j+1)(2j^\prime +1)\over k} \right] 
\chi^{N=2}_{j^\prime m' s^\prime}(\tau).}
Combining this with \modtheta, \modthetaaa\ one finds that 
\eqn\chimonmod{\chi^{\rm monopole}_{js} (-{1\over\tau})
= {1\over\sqrt{2k}}\sum_{j^\prime s^\prime}\CC_{ss'}
\sin\left[ {\pi (2j+1)(2j^\prime +1)\over k} \right] 
\chi^{\rm monopole}_{j's'}(\tau)}
in agreement with the transformation of the untwisted
characters \chisusmod.

Therefore, just like in the bosonic case, in any SCFT with an
$\widehat{SU(2)}$ symmetry we can replace the left-moving 
supersymmetric affine $SU(2)$ sector with a monopole SCFT
without spoiling modular invariance. 

\listrefs
\bye